\documentclass[]{aa}  

\usepackage{graphicx}
\usepackage{amsmath}  
\usepackage{mathrsfs} 
\usepackage{txfonts}
\usepackage{placeins}
\usepackage{float}

\usepackage{color}  
\usepackage{xcolor}
\usepackage{siunitx} 
\usepackage[breaklinks]{hyperref}
\hypersetup{
  colorlinks=true,
    linkcolor=blue,
    citecolor=blue,
    urlcolor=blue}
\usepackage{color}
\usepackage{natbib,twoopt}
\usepackage[hyphenbreaks]{breakurl}
\bibpunct{(}{)}{;}{a}{}{,}            
\definecolor{cobalt}{rgb}{0.06, 0.2, 0.65}
\hypersetup{
  citecolor=cobalt,
  linkcolor=[rgb]{0.8, 0.2, 1.0},
  urlcolor=cobalt,
}
\makeatletter
  \newcommandtwoopt{\citeads}[3][][]{\href{http://adsabs.harvard.edu/abs/#3}%
    {\def\hyper@linkstart##1##2{}%
     \let\hyper@linkend\@empty\citealp[#1][#2]{#3}}}
  \newcommandtwoopt{\citepads}[3][][]{\href{http://adsabs.harvard.edu/abs/#3}%
    {\def\hyper@linkstart##1##2{}%
     \let\hyper@linkend\@empty\citep[#1][#2]{#3}}}
  \newcommandtwoopt{\citetads}[3][][]{\href{http://adsabs.harvard.edu/abs/#3}%
    {\def\hyper@linkstart##1##2{}%
     \let\hyper@linkend\@empty\citet[#1][#2]{#3}}}
  \newcommandtwoopt{\citeyearads}[3][][]%
    {\href{http://adsabs.harvard.edu/abs/#3}
    {\def\hyper@linkstart##1##2{}%
     \let\hyper@linkend\@empty\citeyear[#1][#2]{#3}}}
\makeatother
\newcommand{\feh}{\ensuremath{[\textup{Fe}/\textup{H}]}\xspace}
\newcommand{\stellarmass}{\ensuremath{M}\xspace}
\newcommand{\initialy}{\ensuremath{Y_{\textup{ini}}\xspace}}

\newcommand{\alphamlt}{\ensuremath{\alpha_{\textup{mlt}}}\xspace}

\newcommand{\lum}{\ensuremath{L}\xspace}
\newcommand{\teff}{\ensuremath{T_{\text{eff}}}\xspace}
\newcommand{\surffeh}{\ensuremath{\feh}\xspace}
\newcommand{\age}{\ensuremath{\tau}\xspace}

\newcommand{\Pitchfork}{\texttt{Pitchfork}\xspace}
\newcommand{\UltraNest}{\texttt{UltraNest}\xspace}
\newcommand{\MESA}{\texttt{MESA}\xspace}

\newcommand{\KBa}{\ensuremath{a}\xspace}
\newcommand{\KBb}{\ensuremath{b}\xspace}

\begin{document}

   \title{Characterizing six seismic solar analogs observed by \emph{Kepler}, K2, and HERMES}


   \author{R.~A.~García\inst{1}
          \and
          S.~Mathur\inst{2,3}
          \and
          G.~T.~Hookway\inst{4}
          \and
          D.~Godoy-Rivera\inst{2,3}    
          \and
          T.~Masseron\inst{2,3}
          \and
           J. Bétrisey\inst{5}
          \and
          G. Buldgen\inst{6}
          \and
          C.~Lindsay\inst{7}
          \and
          T.~S. Metcalfe\inst{8}
          \and
          O.~J.~Scutt\inst{4} 
          \and          
          A.~Stokholm\inst{4}
          \and
          P.~G.~Beck\inst{2,3}
          \and
          O.~Benomar\inst{9,10,11}
          \and
          G.~R.~Davies\inst{4}  
          \and
          A.~Jiménez\inst{2,3}
          \and
          J.~Merc\inst{2,12}
          \and 
          M.~B.~Nielsen\inst{4}
          \and
          E.~Panetier\inst{13}
          \and
          F.~Pérez Hernández\inst{2,3}
          \and 
          L.~Borg\inst{1}
          \and
          S.~N.~Breton\inst{14}
          \and
          L.~Debacker\inst{1,15} 
          \and
          A.~Escorza\inst{2,3}
          \and
          D.~H.~Grossmann\inst{2,3}
          \and
          A.~Hamy\inst{1,16}
          \and
          B.~Liagre\inst{17,1,2,3}
          \and
          M.~N.~Lund\inst{18}
          \and
          S.~Mathis\inst{1}
          \and
          D.B.~Palakkatharappil\inst{1}
          \and
          A.~R.~G.~Santos\inst{1}
          \and
          V.~Delsanti\inst{1,2,3,16}
          \and
          L.~González-Cuesta\inst{2,3}          
          \and
          V.~Fox\inst{1,16}
          \and
          N.~Proust\inst{1,16}
          }
          
   \institute{Universit\'e Paris-Saclay, Universit\'e Paris Cit\'e, CEA, CNRS, AIM, 91191, Gif-sur-Yvette, France
   \email{rgarcia@cea.fr}
   \and
    Instituto de Astrof\'isica de Canarias (IAC), E-38205 La Laguna, Tenerife, Spain
    \and
    Universidad de La Laguna (ULL), Departamento de Astrof\'isica, E-38206 La Laguna, Tenerife, Spain
    \and   
    School of Physics and Astronomy, University of Birmingham, Birmingham B15 2TT, UK
    \and 
    Department of Physics and Astronomy, Uppsala University, Box 516, SE-751 20 Uppsala, Sweden
    \and 
    STAR Institute, Université de Liège, Liège, Belgium
    \and
    Department of Astronomy, Yale University, PO Box 208101, New Haven, CT 06520-8101, USA
    \and
    Center for Solar-Stellar Connections, WDRC, 9020 Brumm Trail, Golden, CO 80403, USA
    \and 
    Department of Astronomical Science, School of Physical Sciences, SOKENDAI,
2-21-1 Osawa, Mitaka, Tokyo, 181-8588, Japan
    \and
    Solar Science Observatory, National Astronomical Observatory of Japan,
2-21-1 Osawa, Mitaka, Tokyo, 181-8588, Japan
    \and
    Center for Space Science, New York University Abu Dhabi,
PO Box 129188, Abu Dhabi, UAE
    \and
    Astronomical Institute of Charles University, V Hole\v{s}ovi\v{c}k{\'a}ch 2, Prague, 18000, Czech Republic
    \and
    Universit\'e Paris Cit\'e, Universit\'e Paris-Saclay, CEA, CNRS, AIM, 91191, Gif-sur-Yvette, France 
    \and
    INAF – Osservatorio Astrofisico di Catania, Via S. Sofia, 78, 95123 Catania, Italy
    \and    
    IMT Atlantique, F-29238 Brest, France
    \and
    Ecole Centrale-Supelec, Universit\'e Paris-Saclay, 91190 Gif-sur-Yvette, France
    \and
    ENS Paris-Saclay, Université Paris-Saclay, 91190, Gif-sur-Yvette, France 
    \and
    Stellar Astrophysics Centre, Department of Physics and Astronomy, Aarhus University, Ny Munkegade 120, DK-8000 Aarhus C, Denmark
}
   \date{Received ; accepted}

 
  \abstract
 {
Solar analogs — stars that closely match the fundamental properties of the Sun — provide key benchmarks for testing stellar structure and evolution across different ages and activity levels. Their detailed characterization helps place the Sun in context within the broader population of solar-like stars. This study presents the characterization of six seismic solar analogs observed by the NASA \emph{Kepler} and K2 missions. Combining asteroseismic constraints from space-based photometry with high-resolution spectroscopy and \textit{Gaia} astrometry, we derived their fundamental parameters and assessed their resemblance to the Sun. Global seismic properties and individual oscillation modes were extracted from the photometric light curves, while atmospheric parameters were obtained from data collected by the HERMES spectrograph at the Mercator telescope. Stellar modeling using seven independent stellar evolution codes yielded consistent masses, radii, and ages. These stars have masses between 0.91 and 1.04~$\mathrm{M}_\odot$, radii between 0.95 and 1.08~$\mathrm{R}_\odot$, and ages from about 1.8 to 9.1~Gyr, with typical systematic uncertainties of ±0.02~$\mathrm{M}_\odot$, ±0.01~$\mathrm{R}_\odot$, and ±0.7~Gyr, respectively. 
One star, EPIC~206064678, exhibits properties very similar to those of the Sun, with $M = 1.016 \pm 0.033\,\mathrm{M}_\odot$, $R = 0.990 \pm 0.011\,\mathrm{R}_\odot$, and an age of $5.40 \pm 0.12$\,Gyr. 
It can therefore be considered a close solar twin, although it is slightly older and more metal-rich ($0.25 \pm 0.07$\,dex).
Four targets display binarity signatures and all exhibit very low chromospheric activity. This work broadens the sample of well-characterized seismic solar analogs towards a larger sample of metallicities and ages, providing new references for comparative stellar studies and future asteroseismic investigations. }

   \maketitle
   \nolinenumbers 
%

\section{Introduction}

The Sun is the cornerstone of stellar astrophysics, providing an unparalleled benchmark for the study of stars because of its proximity and the extraordinary detail with which it can be characterized. Its unique status enables investigations of stellar structure and dynamics with a precision unattainable for any other star. However, solar observations provide only a snapshot in the 4.5\,Gyr evolution of our star, leaving fundamental questions unresolved: to what extent is the Sun a typical $\sim$1~$\mathrm{M}_\odot$ solar-like star? What was its past, and what will be its future?

For decades, astronomers have searched for stars with physical properties closely matching those of the Sun, leading to the concepts of solar twins \citep[e.g.][]{1981A&A....94....1C,1989A&A...211..324C} and solar analogs \citep[e.g.][]{1978A&A....63..383H,1996A&ARv...7..243C}. Solar twins are stars whose effective temperature ($T_{\mathrm{eff}}$), surface gravity ($\log g$), metallicity ([Fe/H]), micro-turbulence, photometric properties, chemical composition, age, luminosity, rotation, and magnetic fields are indistinguishable from those of the Sun within observational uncertainties. Because of these stringent criteria, true solar twins are rare, with only about a hundred identified despite extensive observational campaigns \citep[e.g.][and references therein]{2021MNRAS.504.1873Y}. Their close spectroscopic similarity to the Sun nevertheless makes them powerful laboratories for differential abundance analyses, tests of stellar evolution models, and calibrations of stellar parameters. Notable examples include 18~Sco \citep{2012A&A...544A.106B,2018A&A...619A.172B}, HIP~102152, and CoRoT ID~102684698 \citep{2013ApJ...771L..31D}.

Solar analogs form a broader class of Sun-like stars that do not need to reproduce all solar properties. They are typically defined within a mass range of $0.9 \leq M/\mathrm{M_\odot} \leq 1.1$, without constraints on age, allowing a wider diversity in stellar parameters. This flexibility enables solar analogs to represent different evolutionary stages of Sun-like stars and to span a range of rotation rates, magnetic activity levels, and chemical compositions \citep[e.g.][]{2014ApJ...790L..23D,
2020ApJ...898..173D,2016A&A...596A..31S,2016ApJ...826L...2M,2022ApJ...933L..17M,Beck2017,2017EPJWC.16001010G,2025ApJ...983L..31C,2025ApJ...983...70G}. Their study is therefore crucial for reconstructing a ``Sun in Time'' sequence and for exploring how stellar evolution depends on parameters such as metallicity. Examples include KIC~10644253 \citep{2016A&A...589A.118S}, a younger and more active solar analog, KIC~8006161 \citep{2018ApJ...852...46K}, a younger and more metal-rich star, and the binary system 16~Cyg~A/B \citep{2012ApJ...748L..10M,2015MNRAS.446.2959D}, which represents an older counterpart.

A major advance in the characterization of solar-like stars has come from asteroseismology—the study of stellar oscillations. Space missions such as the Convection, Rotation, and planetary Transits mission \citep[CoRoT,][]{2006cosp...36.3749B,2009A&A...506..411A}, \emph{Kepler}/K2 \citep{2009Sci...325..709B,2014PASP..126..398H}, and more recently the Transiting Exoplanet Survey Satellite \citep[TESS,][]{2014SPIE.9143E..20R} have enabled the detection of solar-like oscillations in hundreds of main-sequence stars. These observations provide precise measurements of stellar masses, radii, and ages \citep{2009A&A...506...51B,2011A&A...530A..97B,2017ApJ...835..172L,2022A&A...657A..31M}, surpassing what is achievable from the ground \citep[e.g.][]{2010ApJ...713..935B}. When combined with high-resolution spectroscopy, asteroseismic analyses yield the most accurate stellar parameters currently attainable \citep{2014A&A...569A..21L,2025A&A...696A..42G}. They have also led to the identification of seismic solar analogs, defined as stars whose global seismic parameters—the frequency of maximum power, $\nu_\mathrm{max}$, and the large frequency separation, $\Delta\nu$ \citep[e.g.][]{2019LRSP...16....4G}—are within 10\% of the solar reference values. These stars offer the opportunity to probe the internal structure of Sun-like stars with high precision, although they are not necessarily solar analogs in the classical spectroscopic definition.

Expanding the sample of seismic solar analogs is therefore essential for placing the Sun in a broader stellar context, yet only a few such stars have been characterized \citep[e.g.,][]{2016A&A...589A.118S,2018ApJ...852...46K,2022AJ....163...79H}. In this work, we report a combined seismic and spectroscopic analysis of six additional seismic solar analogs observed by \emph{Kepler}/K2, HERMES \citep[High-Efficiency and High-Resolution Mercator Echelle Spectrograph;][]{raskin2011}, and \textit{Gaia} \citep{2016A&A...595A...1G}. The layout of the paper is as follows. Section~\ref{sec:obs} describes the \emph{Kepler} and K2 observations. Sections~\ref{sec:data}, \ref{sec:binaries}, and \ref{sec:spectroscopic_analysis} present the seismic analysis, multiplicity, and spectroscopic analysis. Stellar structure and evolutionary models are presented in Section~\ref{sec:modeling}. Finally, Section~\ref{sec:discussion} discusses the results and summarizes our conclusions.

\section{Observations}
\label{sec:obs}
From a pre-analysis of \emph{Kepler} and K2 observations, we selected six seismic solar analogs with global seismic parameters, extracted from the A2Z pipeline \citep{2010A&A...511A..46M}, lying within 10$\%$ of the solar references (3050 and 135 $\mu$Hz  for $\nu_{\mathrm{max}}$ and $\Delta \nu$ respectively). 
The first star, KIC~3241581, was observed for one month in short cadence \citep{2010ApJ...713L.160G} by  \emph{Kepler}  \citep{2022A&A...657A..31M}, and was also extensively spectroscopically characterized with HERMES \citep{Beck2016} on the Mercator telescope.  We used \emph{Kepler} Simple Aperture Photometry flux (SAPflux) light curves corrected by the \emph{Kepler} Asteroseismic Data Analysis and Calibration Software (KADACS\footnote{These data products are known as KEPSEISMIC and are available on MAST: \url{http://dx.doi.org/10.17909/t9-mrpw-gc07}}) described in \citet{2011MNRAS.414L...6G}, filtered with a high-pass filter with a cutoff at 3 days, and gap-filled using a multiscale discrete cosine transform following inpainting techniques \citep{2014A&A...568A..10G,2015A&A...574A..18P}.

The remaining five targets were observed by the K2 mission in short-cadence mode. Three of them, namely EPIC~206064678, EPIC~206245055, and EPIC~206371648, were observed during Campaign~C3\footnote{\url{https://keplergo.github.io/KeplerScienceWebsite/k2-approved-programs.html}}, while EPIC~212624487 was observed during Campaign~C6. The fifth target, EPIC~212708252, was observed in two campaigns, C6 and C17, and was analyzed following the same methodology as described in \citet{2023A&A...674A.106G}.
A search for signatures of a possible magnetic activity cycle in EPIC~212708252 was carried out by \citet{2026MNRAS.546ag092B}, but no statistically significant evidence was found. In addition, EPIC~212708252 and EPIC~212624487 were previously analyzed by \citet{2024A&A...688A..13L}, who derived their global seismic parameters, and by \citet{2025MNRAS.544.3247H}, who characterized their individual oscillation modes.
For all \emph{K2} targets, we adopted light curves from the Ecliptic Plane Input Catalog Variability Extraction and Removal for Exoplanet Science Targets pipeline \citep[EVEREST\footnote{\url{https://stdatu.stsci.edu/prepds/everest/}};][]{2016AJ....152..100L,2018AJ....156...99L}. The light curves were subsequently gap-filled using the same procedure applied to the \emph{Kepler} target.

HERMES spectroscopy was obtained for all the targets. To reach high signal-to-noise ratios, the observations were split into multiple integrations, typically spanning $\sim$400 days. This also allows for the search of binarity signatures (see Sect.~\ref{sec:spectroscopic_analysis}).

\section{Data Analysis}
\label{sec:data}

Two complementary seismic analyses are done for each target. First, we extracted the global seismic parameters that describe the overall oscillation pattern of the star. Second, we characterized the individual oscillation modes through detailed peak-bagging procedures using two independent fitting codes. 

\begin{table*}[!htb]
\caption{Seismic and spectroscopic parameters of the six seismic solar analogs.\label{tab:seismic-spectro}}
\tabcolsep=2pt
\centering
\small
\begin{tabular}{r|rr|rrrr|rrrr}
\hline \hline
\multicolumn{1}{c|}{Identifier} & 
\multicolumn{2}{c|}{A2Z} & 
\multicolumn{4}{c|}{Universal Pattern with \texttt{apollinaire}} &
\multicolumn{4}{c}{Spectroscopic Parameters} \\
\hline
\multicolumn{1}{c|}{} &
\multicolumn{1}{c}{$\nu_\mathrm{max}$} & 
\multicolumn{1}{c|}{$\Delta\nu$} & 
\multicolumn{1}{c}{$\nu_\mathrm{max}$} &
\multicolumn{1}{c}{$\Delta\nu$} &
\multicolumn{1}{c}{$\epsilon$} &
\multicolumn{1}{c|}{$n_{\mathrm{fit}}$} &
\multicolumn{1}{c}{$T_\mathrm{eff}$} & 
\multicolumn{1}{c}{$\log g$} & 
\multicolumn{1}{c}{[Fe/H]} & 
\multicolumn{1}{c}{$v\sin i$} \\
\multicolumn{1}{c|}{} &
\multicolumn{1}{c}{[$\mu$Hz]} &
\multicolumn{1}{c|}{[$\mu$Hz]} &
\multicolumn{1}{c}{[$\mu$Hz]} &
\multicolumn{1}{c}{[$\mu$Hz]} &
\multicolumn{1}{c}{[--]} &
\multicolumn{1}{c|}{[--]} &
\multicolumn{1}{c}{[K]} &
\multicolumn{1}{c}{[dex]} &
\multicolumn{1}{c}{[dex]} &
\multicolumn{1}{c}{[km/s]} \\
\hline    
EPIC~206064678 (a)& 3272\,$\pm$\,93 & 138.06\,$\pm$\,3.97 & 3146\,$\pm$\,118 & 137.80\,$\pm$\,0.18 & 1.564\,$\pm$\,0.029 & 5 & 5476\,$\pm$\,55 & 4.41\,$\pm$\,0.11 & 0.25\,$\pm$\,0.07 & 2.3\,$^{+1.5}_{-2.1}$ \\
EPIC~206245055 (b)& 3187\,$\pm$\,76 & 138.69\,$\pm$\,3.71 & 3129\,$\pm$\,52 & 138.78\,$\pm$\,0.18 & 1.485\,$\pm$\,0.029 & 5 & 5841\,$\pm$\,54 & 4.47\,$\pm$\,0.17 & -0.39\,$\pm$\,0.09 & 3.3\,$^{+1.5}_{-2.1}$  \\
EPIC~206371648 (c)& 2972\,$\pm$\,70 & 139.03\,$\pm$\,4.82 & 2993\,$\pm$\,63 & 136.02\,$\pm$\,0.18 & 1.512\,$\pm$\,0.026 & 9 & 5573\,$\pm$\,43 & 4.37\,$\pm$\,0.17 & -0.26\,$\pm$\,0.06 & 4.7\,$^{+1.4}_{-2.0}$ \\
EPIC~212624487 (d)& 3109\,$\pm$\,150 & 130.78\,$\pm$\,3.70 & 3072\,$\pm$\,21 & 132.75\,$\pm$\,0.10 & 1.387\,$\pm$\,0.017 & 9 & 6038\,$\pm$\,27 & 4.50\,$\pm$\,0.15 & -0.16\,$\pm$\,0.09 & 4.7\,$^{+1.8}_{-2.5}$ \\
EPIC~212708252 (e)& 2930\,$\pm$\,84 & 135.11\,$\pm$\,4.09 & 2913\,$\pm$\,20 & 133.40\,$\pm$\,0.05 & 1.542\,$\pm$\,0.008 & 9 & 5579\,$\pm$\,43 & 4.42\,$\pm$\,0.13 & -0.09\,$\pm$\,0.07 & 2.2\,$^{+1.5}_{-2.1}$ \\
KIC\,3241581 (f)& 2751\,$\pm$\,122 & 122.90\,$\pm$\,1.60 & 2828\,$\pm$\,171 & 122.84\,$\pm$\,0.88 & 1.435\,$\pm$\,0.276 & 5 & 5730\,$\pm$\,50 & 4.39\,$\pm$\,0.13 & 0.25\,$\pm$\,0.06 & 3.5\,$^{+1.6}_{-2.2}$ \\
\hline
\end{tabular}

\label{tab:Seismic}
\flushleft {\bf Notes.} The star’s identifier is the Ecliptic Plane Input Catalog \citep[EPIC,][]{2016ApJS..224....2H} or the \emph{Kepler} Input Catalog \citep[KIC,][]{2011AJ....142..112B}. The frequency of maximum power $\nu_{\mathrm{max}}$ and large separation $\Delta\nu$ are given from the A2Z pipeline. The Universal Pattern parameters computed with the \texttt{apollinaire} code include $\nu_{\mathrm{max}}$, $\Delta\nu$, the phase offset $\epsilon$, and the number of fitted radial orders $n_{\mathrm{fit}}$. Spectroscopic parameters include effective temperature $T_{\mathrm{eff}}$, surface gravity $\log g$, metallicity [Fe/H], and projected rotational velocity $v \sin i$, all with their respective uncertainties.
\end{table*}

\subsection{Extracting global seismic parameters}

We analyzed the power spectral densities (PSDs) of the six targets using A2Z. This pipeline identifies the regular spacing of oscillation modes in the PSD by computing the power spectrum of the power spectrum, from which $\Delta\nu$ is determined. The background is modeled as the sum of several components: a power law to account for magnetic activity, two Harvey-like profiles \citep{1985ESASP.235..199H} representing different convective timescales, and a constant white-noise term to take into account the contribution of the photon noise. $\nu_{\mathrm{max}}$ is then derived by fitting a Gaussian envelope to the background-corrected PSD. A second inference of the seismic global parameters as well as the phase offset, $\epsilon$, is obtained using the universal pattern module of the \texttt{apollinaire} package \citep[see the details in][]{2022A&A...663A.118B}. 
The results from A2Z and \texttt{apollinaire} are summarized in Table~\ref{tab:seismic-spectro}. As expected, the values of the two global seismic parameters, $\Delta\nu$ and $\nu_{\rm max}$, obtained with both pipelines are consistent within 1 $\sigma$ for all stars.

\subsection{Characterizing individual modes: Peak-bagging} 

To characterize the individual oscillation modes, we applied two independent peak-bagging pipelines: the \texttt{apollinaire} module and \texttt{PBjam} \citep{2021AJ....161...62N, Nielsen_2025}. These tools were used to fit and extract the mode frequencies, amplitudes, and linewidths from the observed PSDs. The following subsections describe the implementation and fitting strategies adopted for each method, and present the resulting frequency lists used for the stellar modeling.

\subsubsection{\texttt{apollinaire}}
\label{Subsect:apollinaire}
The \texttt{apollinaire} module\footnote{\url{https://apollinaire.readthedocs.io/en/latest}} implements an ensemble Markov Chain Monte Carlo (MCMC) approach using the \texttt{emcee} sampler \citep{2013PASP..125..306F}.
Each oscillation mode, identified by its radial order ($n$) and angular degree ($\ell$), is modeled as a symmetric Lorentzian profile characterized by its central frequency ($\nu_{n,\ell}$), linewidth ($\Gamma_{n,\ell}$), and height ($H_{n,\ell}$).
In practice, the parameters sampled by the MCMC procedure are $\nu_{n,\ell}$, $\log A_{n,0}$, $\log \Gamma_{n,\ell}$, and the mode visibility ratios ($V_\ell$). The mode amplitudes, $A_{n,\ell} = V_\ell A_{n,0}$, are related with the other mode properties as $A_{n,\ell} = \sqrt{\pi \Gamma_{n,\ell} H_{n,\ell} / 2}$ \citep[e.g.][]{2017ApJ...835..172L}.
This parameterization ensures uniform prior distributions and facilitates efficient convergence of the MCMC chains.
The likelihood is constructed assuming that the statistical fluctuations of the observed power spectrum about the model follow a $\chi^2$distribution with two degrees of freedom.

\subsubsection{\texttt{PBjam}}

The mode characterization was also carried out using \texttt{PBjam}, following a two-stage approach consisting of (1) mode identification based on asymptotic relations, and (2) a detailed peak-bagging analysis.
During the mode identification stage, we applied a background model to the power spectrum composed of three Harvey-like profiles~\citep{1985ESASP.235..199H} and a white-noise component. This background was then combined with a model of the $\ell = 0, 1,$ and $2$ oscillation modes, each represented as the sum of Lorentzian profiles~\citep{1990ApJ...364..699A}. The mode frequencies were constrained by the asymptotic relation for p modes, in which the parameters $\Delta\nu$, $\nu_{\mathrm{max}}$, $\varepsilon$, and the small separations $\delta\nu_{01}$ and $\delta\nu_{02}$ were treated as free parameters.
The number of radial orders included in the model was varied between 6 and 14, depending on the signal-to-noise ratio (S/N) of the power spectrum. The \texttt{Dynesty} nested sampling algorithm~\citep{Speagle_2020} was used to sample the posterior probability distribution of the model parameters, and thereby determine the most probable mode frequencies. The prior distribution was constructed from a large dataset of thousands of previously analyzed stars and stellar models, allowing empirical knowledge of stellar oscillations to inform the mode identification posterior.

The results from the mode identification were then passed to the second stage of \texttt{PBjam}, corresponding to the detailed peak bagging. In this step, each oscillation mode was fitted individually, allowing for rotational splitting and acoustic glitches not captured in the asymptotic approximation. Normal priors were applied to the mode frequencies, logarithmic heights, and logarithmic linewidths, with prior widths of 3\% of $\Delta\nu$ for the frequencies and 50\% of the mode identification values for the heights and widths. The fitting procedure was performed using the \texttt{emcee} MCMC sampler.
Finally, to validate the results, we computed the ratio between the median absolute deviation (MAD) of the posterior and prior distributions for each mode frequency. Modes with a ratio below 50\% were deemed to contain new information from the observed light curve, and only these modes were retained.

\subsubsection{Consolidated peak-bagging results}

The peak-bagging results for each star are presented in Appendix~\ref{App:extraction}. For each star, there is a table listing the radial order $n$ and the angular degree $\ell$ of the detected modes, along with their frequencies and symmetrized uncertainties derived from the two fitting methods. We compared the agreement between the two pipelines, and found their frequencies agreed within 1 $\sigma$ for 74.5\,\% of the modes across all targets. 

\section{Binarity of the targets}
\label{sec:binaries}

About 50\,$\%$ of main-sequence, solar-like stars with spectral types from late F to early K are expected to be in multiple systems \citep[e.g.][]{2006ApJ...640L..63L,MoeStefano2017}, whereas this fraction decreases to about 30$\%$ for solar-like pulsating stars \citep[e.g.][]{2026A&A...707A.298B}. Stars in close binary systems may undergo tidal and magnetic interactions that can affect their oscillation modes \citep[e.g.][]{2010Sci...329.1032G,2011ApJ...732L...5C,2019FrASS...6...46M,2014ApJ...785....5G,2020A&A...639A..63G,gehan24}. It is therefore important to assess the multiplicity and the orbital separations between the stars for our stellar sample. Indeed, since we will use stellar evolution codes that assume single-star evolution, it is important to know whether the inferred stellar parameters could be biased.

Four out of the six target stars are reported to be members of binary or multiple stellar systems (see Table~\ref{tab:Spectro}). A companion to EPIC\,206064678 was detected via speckle interferometry at a separation of 0.8 arcseconds and a position angle of approximately 300 degrees. The companion is fainter by 3.2 magnitudes in the I-band \citep{2010AJ....139..205H, 2020AJ....160....7T}. This binary is also resolved by \textit{Gaia}, which reports an angular separation of 0.814 arcseconds and a consistent position angle. Although the companion lacks parallax and proper motion measurements in \textit{Gaia} DR3, the relatively high proper motion of EPIC\,206064678, combined with the reasonably long time baseline between the interferometric and \textit{Gaia} observations (>8 years), makes a chance alignment with a background star highly unlikely. If the two objects were not co-moving, a significant shift in relative position would be expected. Assuming a distance of 57.6 pc (from the inverse of the \textit{Gaia} DR3 parallax), the projected physical separation is estimated to be $\sim$47 au, suggesting an orbital period of several hundred years. However, the current time span of interferometric observations is insufficient to derive a reliable orbital solution.

\begin{table*}[!htb]
\tiny

\caption{Photometric, radial-velocity, and astrometric information of the six candidate solar analogs.}
\tabcolsep=2pt
\centering

\begin{tabular}{r|rrrrrr|rrrrr|rl}
\hline \hline
\multicolumn{1}{c|}{Identifier} & 
\multicolumn{1}{c}{G} & 
\multicolumn{1}{c}{N} & 
\multicolumn{1}{c}{ToT} & 
\multicolumn{1}{c}{$\Delta T$} & 
\multicolumn{1}{c}{RV} & 
\multicolumn{1}{c|}{$\Delta$RV} & 
\multicolumn{1}{c}{A(Li)} &
\multicolumn{1}{c}{$S$-index} &
\multicolumn{1}{c}{$\log R'_\mathrm{HK} (B-V)$} &
\multicolumn{1}{c}{$\log R'_\mathrm{HK} (T_{\mathrm{eff}})$} &
\multicolumn{1}{c|}{$\log R'_\mathrm{IRT}$} &
\multicolumn{1}{c}{RUWE} &
\multicolumn{1}{c}{Multiple} \\
& 
\multicolumn{1}{c}{[mag]} & 
& 
\multicolumn{1}{c}{[hrs]} & 
\multicolumn{1}{c}{[d]} & 
\multicolumn{1}{c}{[km/s]} & 
\multicolumn{1}{c|}{[km/s]} & 
\multicolumn{1}{c}{} &
\multicolumn{1}{c}{} &
\multicolumn{1}{c}{} &
\multicolumn{1}{c}{} &
\multicolumn{1}{c|}{} &
\multicolumn{1}{c}{} &
\multicolumn{1}{c}{} \\
\hline
EPIC~206064678 (a)& 8.73 &4&1.7&422&19.15&0.04& $<$1.0 & 0.2093\,$\pm$\,0.0230 & $-5.00\pm0.07$ & $-4.93\pm0.09$& $-6.554 \pm 0.074$ & 1.1 & Binary \\
EPIC~206245055 (b)& 8.60 &4&0.8&422&-33.0& 0.01&2.0 $\pm$0.05 & 0.1633\,$\pm$\,0.0230 & $-5.02\pm0.10$ & $-5.03\pm0.16$& $-5.363 \pm 0.007$ & 0.8 &  \\
EPIC~206371648 (c)& 8.65 &5&1.2&416&-23.71& 0.07& $<$1.0 & 0.1909\,$\pm$\,0.0230 & $-4.97\pm0.06$ & $-4.98\pm0.11$&  &   9.1 &  Triple\\
EPIC~212624487 (d)& 7.55 &6&1.1&426&-43.16& 0.10&2.55 $\pm$ 0.05 & 0.1817\,$\pm$\,0.0115 & $-4.98\pm0.05$ &$-4.85\pm0.06$& $-5.654 \pm 0.008$ & 1.0 &  \\
EPIC~212708252 (e) & 7.89 &7&1.6&3017&14.2 &2.49 & $<$ 0.8 & 0.1771\,$\pm$\,0.0069 & $-5.01\pm0.03$ &$-5.05\pm0.04$& $-5.622 \pm 0.006$ & 1.3 & Binary \\
KIC\,3241581   (f) & 10.15  & 40&13.9&2303&-31.64&2.08& $<$ 1.0 & 0.1770\,$\pm$\,0.0920 & $-4.93\pm0.03$ &$-4.99\pm0.53$&  & 6.4 & Binary \\
\hline
\end{tabular}
\label{tab:Spectro}
\flushleft {\bf Notes.} The columns are the star’s identifier EPIC and KIC, apparent magnitude in the \textit{Gaia} G filter, number of spectra (N), total accumulated time on target (ToT), the time base $\Delta T$ covered by the observations, the mean radial velocity (RV), the difference between the positive and negative extrema of the measured RV values $\Delta$RV, the abundance (or upper limit) of the lithium abundance A(Li), the $S-$, $\log R'_{\mathrm{HK}} (B-V)$, $\log R'_{\mathrm{HK}} (T_\mathrm{eff})$, and $\log R'_{\mathrm{IRT}}$ indexes, the \textit{Gaia} DR3 RUWE value, and a last column explaining if the star belongs to a confirmed binary or a triple system. 
\end{table*}

EPIC\,206371648 is listed in the 9th Catalog of Spectroscopic Binary Orbits \citep[SB9;][]{2004A&A...424..727P} under entries SBC9 3701 and SBC9 3702. \citet{Tokovinin2018} identified EPIC\, 206371648 as a hierarchical triple system, consisting of an inner spectroscopic binary with a period of 111.1 days and an outer visual companion with an orbital period of 18,440\,$\pm$\,1,576\,days ($\sim$50.49\,$\pm$\,4.3\,years) and eccentricity $e = 0.451\,\pm$\,0.044. These parameters were derived from extensive speckle interferometric data. \citet{Videla2022} derived a mass ratio of $\sim$0.52--0.96 between the inner pair and the outer companion, depending on the input data. The system shows a high Renormalized Unit Weight Error (RUWE $\simeq$ 9.1) in \textit{Gaia} DR3 (RUWE higher than $\sim$1.25 indicates either multiplicity or a poor astrometric solution; \citealt{2022MNRAS.513.2437P}), yet no solution is present in \textit{Gaia} DR3 Non-Single Star (NSS) catalog \citep{Arenou2023}, in either the two-body orbit tables (TBO) or non-linear proper motion tables (NLAC). The star is also flagged as an SB2 candidate in \citet{2023ApJS..266...18Z}, based on Large Sky Area Multi-ObjectFiber Spectroscopic Telescope \citep[LAMOST,][]{2012RAA....12..723Z,2020ApJS..251...15Z} spectra and machine learning classification. However, the apparent SB2 signature may result from the combined light of the outer companion and one of the inner components, rather than an actual double-lined spectroscopic nature of the inner pair. The asymmetry in the cross-correlation function (CCF) observed by \citet{Tokovinin2018} at some epochs was explained in this way. Although the absorption lines of the secondary component are hardly distinguishable in the spectrum, we have been able to estimate the maximum contribution of the secondary component to the spectral analysis of the primary, assuming that the secondary component is a main-sequence star. We find that the change in the effective temperature is not significant within the measured uncertainties. Only the metallicity and 
$v\sin i$ are affected by the secondary contribution, resulting in an increase in metallicity of up to 0.05 dex and a decrease in velocity of $\sim$2 km/s.
As these values are comparable within 1\,$\sigma$ of the original measurements, we decided to retain them.

EPIC\,212708252 was identified as a long-period spectroscopic binary with $P = 61,223.07^{+2585.17}_{-3060.66}$ days ($\simeq$167 years) based on 25 years of radial velocity monitoring with CORALIE \citep{Barbato2023}. The orbit is highly eccentric ($e\sim$0.73), with a peak-to-peak RV amplitude of $\sim$4 km/s. The RUWE is moderate ($\sim$1.3), and no solution is listed in either TBO or NLAC, as expected for a system with such a long orbital period. However, the system shows a significant proper motion anomaly \citep{2019A&A...623A..72K}, consistent with its binary nature.

Finally, KIC\,3241581 was reported to be a binary by \citet{Beck2016}. The system appears in the \textit{Gaia} DR3 NSS catalog with an acceleration solution and a relatively high RUWE$\sim$6. This target was continuously monitored with the HERMES spectrograph \citep{raskin2011}, mounted on the 1.2-meter Mercator Telescope at the Roque de los Muchachos Observatory in La Palma, Canary Islands, Spain. 
These 40 observations over 2,303 days have enabled further constraints on its orbital parameters. Using the {\tt radial} code\footnote{\url{https://github.com/ladsantos/radial}}, we analyzed the RVs from HERMES monitoring to refine the orbital elements. The resulting parameters are listed in Table~\ref{tab:orbit_3241581}, and the orbital fit is shown in Fig.~\ref{fig:orbit_3241581}.

From this analysis we conclude that the periods of the companions are long. As a consequence the oscillating star can be treated as a single star without tidal interaction that would affect the oscillation mode pattern \citep{2014ApJ...785....5G, Beck2024}.

\section{Atmospheric parameters}
\label{sec:spectroscopic_analysis}

To ensure a consistent dataset of fundamental parameters, high-resolution spectra were collected for each target using the HERMES spectrograph. It has a spectral resolution of R$\sim$85,000 over a wavelength range spanning 375–900 nm. The wavelength calibration was achieved using a thorium-argon-neon (ThArNe) emission lamp. 

Observations were carried out between June 2019 and August 2020, except for EPIC~212708252 and KIC\,3241581 which have been followed up with the same spectrograph since 2012, thus covering a longer time range as detailed in Tables~\ref{tab:Seismic} and \ref{tab:Spectro}.
The spectroscopic data were reduced using the instrument's dedicated pipeline \citep{raskin2011}. These observations are part of a large follow up program focused on seismic solar-analog stars, described in detail in \cite{Beck2016,Beck2017}. The radial velocity (RV) for individual spectra was calculated via the CCF with a standard G2-mask across the wavelength range of 478–653 nm, as implemented in the HERMES data reduction toolbox. For this observational setup, the spectrograph's night-to-night RV stability at the 3$\sigma$ level is approximately 300 m/s, a conservative threshold often used to identify binary systems. To ensure RV stability, several ThArNe reference frames were taken throughout the night between science exposures.

The combined stellar spectra were analyzed using the 1D/LTE \texttt{BACCHUS} code \citep{2016ascl.soft05004M,Hayes2022} with MARCS model atmospheres \citep{2008A&A...486..951G} and atomic and molecular linelists from \citet{Heiter2021}. The effective temperature was determined by requiring no trend between the abundances of \ion{Fe}{I} lines and their excitation potentials. The surface gravity was obtained from ionization balance between \ion{Fe}{I} and \ion{Fe}{II}, while the microturbulent velocity was derived by removing trends between Fe-line abundances and equivalent widths. The metallicity corresponds to the mean abundance of the \ion{Fe}{I} lines. The code simultaneously fits the Fe-line profiles with a mean line-broadening parameter. The projected rotational velocity ($v\sin i$) was estimated assuming that this broadening results from the quadratic sum of instrumental resolution ($R=85\,000$), macroturbulence following \citet{Doyle2014} for G–K main-sequence stars, and rotational broadening. For slow rotators, $v\sin i$ derived in this way remains uncertain because of degeneracies with macroturbulence, possible instrumental resolution variations, and uncertainties in line parameters. In addition, the spectral energy distribution of each star was analyzed with the VOSA tool \citep{Bayo2008}. The resulting temperatures are fully consistent with those from the spectroscopic analysis. As a final check, the wings of the H$_\alpha$ and Mg\,\textsc{i} triplet lines, which are sensitive to temperature and pressure, were visually inspected. Once the stellar parameters were fixed, lithium abundances (or upper limits) were derived from the 6707\,\AA\ line. The resulting stellar properties are listed in Table~\ref{tab:Spectro}.

We note that our stars are not solar analogs in the classical spectroscopic term \citep[e.g.][]{1996A&ARv...7..243C}. Indeed, the target selection was based purely on the global seismic parameters as explained in Section~\ref{sec:obs}. This corresponds to an extended version of the classical definition of solar analogs as the stars have $T_\mathrm{eff}$ within 300~K of the solar value and [Fe/H] within 0.4~dex, similarly to those in  \citet{2016A&A...596A..31S}. 

We complete the spectroscopic analysis by computing the $S$-index following \citet{Beck2016} (multiplied by 23 to match the Mount Wilson Observatory scale) and the $\log R'_{HK}(B-V)$ index following \citet{Mascareno2015} for each individual exposure of the HERMES spectra. We note that low activity indicators are especially sensitive to systematic measurement errors. Consequently, values below the solar mean activity level ($\sim$ 0.17  for the S-index \citep{2017ApJ...835...25E} and -4.91 for the $\log R'_{HK} (B-V)$ \citep{Mamajek2008}) should be interpreted with caution, particularly when comparing different studies, spectral types, or instruments.

We also calculated the \ion{Ca}{ii} infrared triplet (IRT) index following the approach of \citet{godoyrivera26b}. In brief, the $\log{R^{\prime}_{\mathrm{IRT}}}$ index was computed from the \texttt{activityindex\_espcs} parameter ($\alpha$) in the \texttt{gaiadr3.astrophysical\_parameters} table, which quantifies the excess equivalent width with respect to a reference spectrum at the \ion{Ca}{ii} IRT as observed by the \textit{Gaia} Radial Velocity Spectrometer \citep{2023A&A...674A..30L}. Out of the six targets in our sample, four of them have measured equivalent widths with positive values, where a $\log{R^{\prime}_{\mathrm{IRT}}}$ could be calculated, as well as a signal-to-noise ratio (SNR) of $\alpha/\sigma_{\alpha}>3$. The values of both indices are given in Table~\ref{tab:Spectro}. 

A detailed analysis of KIC~3241581 can be found in \citet{Beck2016}. Here, we obtain results that are highly consistent with our previous results, although with somewhat larger uncertainties. This difference arises because the previous study employed a differential approach with respect to the Sun, which leads to improved accuracy. Nevertheless, for the sake of consistency across our sample, we adopt our own values for the modeling.

For completeness, we searched for stellar parameters of our targets in several spectroscopic surveys: APOGEE (DR17; \citealt{majewski17,abdurrouf22}), LAMOST (DR10; \citealt{cui12,2012RAA....12..723Z}) Medium- and Low-Resolution Spectroscopy (LRS and MRS, respectively), \textit{Gaia}-ESO (v5.1; \citealt{gilmore12,randich13,hourihane23}), GALAH (DR4; \citealt{desilva15,buder24}), and \textit{Gaia} DR3 gspspec \citep{2023A&A...674A...1G,recioblanco23}. We also queried the spectro-photometric catalog of \citet{andrae23}, which used the XGBoost algorithm on the \textit{Gaia} DR3 XP coefficients to derive stellar parameters. 

Out of our six targets, we found two in common with APOGEE, four in common with \textit{Gaia} DR3 gspspec, one in common with LAMOST LRS, and five in common with \citet{andrae23}. No overlap was found with the LAMOST MRS, \textit{Gaia}-ESO, or GALAH surveys. The \citet{andrae23} values are by construction on the APOGEE scale. Following \citet{2025A&A...696A.243G}, for the \textit{Gaia} DR3 gspspec values we only considered the best-quality data and implemented the gravity and metallicity calibration to place them in a common scale with other surveys \citep{recioblanco23}.

\begin{figure}[!t]
\begin{center}
    \includegraphics[width=0.39\textwidth,trim={0cm 0.5cm 0cm 0.5},clip]
    {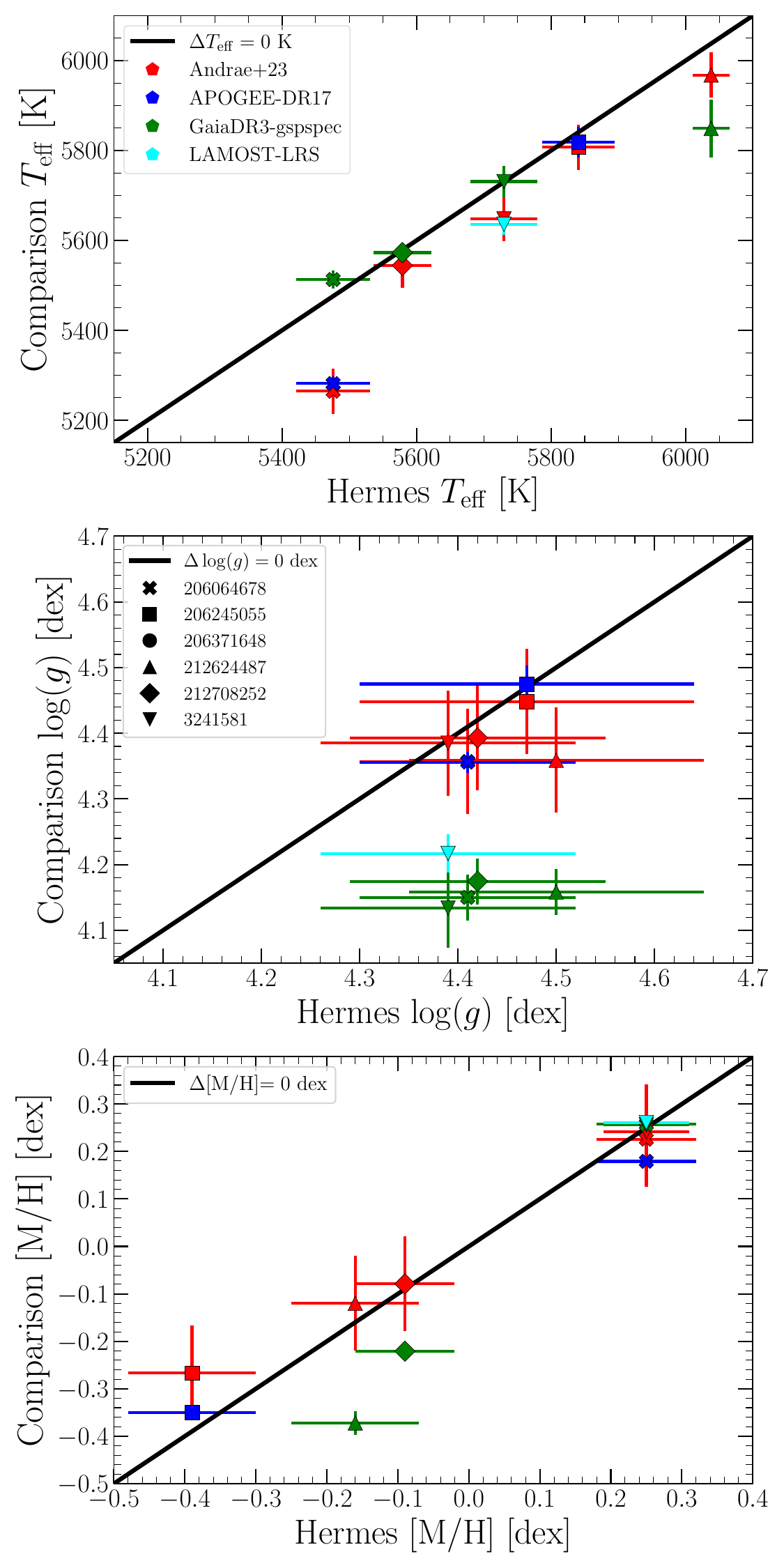}    
\end{center}    
    \caption{Comparison of atmospheric parameters: HERMES vs. spectroscopic surveys. The top, middle, and bottom panels show the comparison of effective temperatures, surface gravities, and metallicities. The HERMES values are on the x-axis, and the spectroscopic surveys are in the y-axis, with \citet{andrae23} in red, APOGEE DR17 in blue, and \textit{Gaia} DR3gspspec in green. Each target is shown in a different symbol, as indicated in the middle panel legend. The \citet{andrae23} values show the better overall agreement, and largest overlap, with HERMES.} 
    \label{fig:ap_spectroscopic_comparison}
\end{figure}

Figure \ref{fig:ap_spectroscopic_comparison} shows the comparison\footnote{Note that, for the purposes of this comparison and Figure \ref{fig:ap_spectroscopic_comparison}, we take [Fe/H] as [M/H] and compare both of them directly.} of the spectroscopic surveys vs. the HERMES values. The five targets in common with \citet{andrae23} (red), and the two targets in common with APOGEE (blue), show a good overall agreement with HERMES. Regarding the four targets in common with \textit{Gaia} DR3 gspspec (green), and the one target in common with LAMOST LRS, while the temperature and metallicity differences are moderate, the surface gravities are substantially under-estimated ($\Delta \log(g) \gtrsim 0.2$ dex) for all of them, hinting to systematic errors.
Because we only have HERMES spectra for all the targets and to ensure homogeneity, we adopted those spectroscopic values throughout this paper.

\section{Luminosity computation}
We calculated luminosities following the approach of \citet{godoyrivera26a}, which we briefly summarize here. First, we calculated absolute magnitudes in the $G$-band using the \textit{Gaia} DR3 astrometric and photometric information \citep{2021A&A...649A...2L,2023A&A...674A...1G}, accounting for the parallax and extinction corrections as in \citet{2025A&A...696A.243G}. Second, we used the \texttt{gaiadr3\_bcg}\footnote{\url{https://gitlab.oca.eu/ordenovic/gaiadr3_bcg}} function \citep{2023A&A...674A..26C}, which interpolates MARCS models \citep{2008A&A...486..951G} taking as input the atmospheric parameters from Table~\ref{tab:seismic-spectro}, and calculated $G$-band bolometric corrections by sampling one thousand Monte Carlo realizations of each target. Third, we combined these absolute magnitudes and bolometric corrections with the solar bolometric magnitude from \citet{2015arXiv151006262M} to compute the luminosities. We validated our values by comparing them with those of \textit{Gaia} DR3 FLAME (final luminosity age mass estimator), which reported luminosities for four (out of six) of our targets, finding that in all four cases both estimates agree at the 3 $\sigma$-level.

As some of our targets are members of multiple systems, we tested the impact of companions on the derived luminosities. For this, we compared the aforementioned \textit{Gaia}-based luminosities (which would include the contribution of companions unresolved by \textit{Gaia}), with the values that can be estimated from the asteroseismic scaling relations (which would only include the brightest oscillating component, $L \propto \nu_{\mathrm{max}}^2 {\Delta\nu}^{-4} T_{\mathrm{eff}}^5$). The comparison between both estimates agree within $1\sigma$ for four out of the six targets, with the remaining two targets agreeing at the 1.3 and 1.5 $\sigma$-levels (EPIC 212708252 and EPIC 206371648, respectively). Thus, we conclude that unresolved companions do not strongly impact the \textit{Gaia}-based luminosities.

\section{Stellar modeling}
\label{sec:modeling}

We modeled the stars to estimate the masses, radii, and ages of the targets. We used several methods and stellar evolution codes, allowing us to measure the systematics.
The input parameters of the optimizing codes are the atmospheric parameters from HERMES ($T_{\rm eff}$ and [Fe/H]), luminosity from \textit{Gaia}, the global seismic parameters ($\Delta \nu$ and $\nu_{\rm max}$), and the individual frequencies of the modes. We note that no correction from the line-of-sight velocity of the stars was applied. Since they are negligible compared to the uncertainties reported, we do not expect any impact as also shown by \citet{2014MNRAS.445L..94D}.

\subsection{IACgrid description}
The IACgrid is composed of models computed with the Modules for Experiments in Stellar Astrophysics \citep[\texttt{MESA}, ][]{2011ApJS..192....3P,2013ApJS..208....4P,2015ApJS..220...15P} stellar evolution code (version 15140). It uses standard input physics with  OPAL opacities \citep{1996ApJ...464..943I} and chemical mixture from \citet{1998SSRv...85..161G}. The masses are in the range 0.8\,$\mathrm{M}_\odot$ to 1.5\,$\mathrm{M}_\odot$ (with a step of 0.01\,$\mathrm{M}_\odot$), initial abundances from -0.3 to 0.4\,dex (with a step of 0.05\,dex), and mixing length parameter $\alpha_{\rm mlt}$ between 1.5 and 2.2 (with a step of 0.05). The mixing length follows the prescription of \citet{Cox&Giuli1968}. The frequencies were computed with the Aarhus adiabatic oscillation package \texttt{ADIPLS} \citep{2008Ap&SS.316..113C}. The primordial helium abundance was taken as $Y_0 = 0.249$. Based on the Galactic chemical evolution model with $\Delta Y/\Delta Z$\, fixed to 1.33, we take an initial solar helium abundance of 0.2744 and initial solar metallicity of 0.0191. These values are consistent with the opacities and chemical abundances used in the models.

The optimization relies on a $\chi^2$ minimization, where the $\chi^2$ is divided into the spectroscopic one, the frequencies one, and the dynamical one. To compare the observed frequencies to the model ones, we apply surface corrections following \citet{2019FrASS...6...41P}. The uncertainties are estimated via Monte Carlo simulations, assuming Gaussian uncertainties on the observables.
More details on the optimization and the computation of the uncertainties on the derived parameters can be found in \citet{2019FrASS...6...41P} and \citet{2023A&A...674A.106G}.

\subsection{BASTA}
The \texttt{BASTA} results are obtained using the Bayesian framework of the Bayesian STellar Algorithm \citep[see][]{aguirre22}, which infers stellar properties from a custom-computed grid of stellar models. This grid, generated with the Garching Stellar Evolution Code (\texttt{garstec}; \citealt{weiss2008}), contains evolutionary tracks constructed under well-established input physics. The equation of state combines contributions from the OPAL group \citep{rogers1996,2002ApJ...576.1064R} and the Mihalas–Hummer–Däppen formulation \citep{mihalas1988,hummer1988,daeppen1988,mihalas1990}. Opacities are taken from OPAL tables \citep{rogers1992,1996ApJ...464..943I} at high temperatures and supplemented with low-temperature data from \citet{Ferguson2005}. Nuclear reaction rates primarily follow the NACRE compilation \citep{angulo1999}, with exceptions for the $^{14}$N(p,$\gamma$)$^{15}$O and $^{12}$C($\alpha$,$\gamma$)$^{16}$O reactions, where the rates from \citet{formicola2004} and \citet{hammer2005}, respectively, were adopted. The models assume the solar composition of \citet{asplund2009}, and convection is treated within the framework of mixing-length theory \citep{bohm1958,kippenhahn2013}.

To construct the grid, the parameter space is sampled using Sobol quasi-random, low-discrepancy sequences \citep{sobol1,sobol2}, ensuring uniform coverage. The grid spans initial iron abundances of $\feh \in (-1.0,0.6)$~dex, stellar masses $\stellarmass \in (0.7, 1.2)$~$\mathrm{M}_{\odot}$, initial helium fractions $\initialy \in (0.24, 0.32)$, mixing-length parameters of $\alphamlt \in (1.6, 2.0)$, with $\alphamlt=1.791$ calibrated to the present-day Sun.
Atomic diffusion is included in the models. For this modeling, we include a prior in initial mass following the \citet{salpeter1955} initial mass function to quantify the expected mass distribution of stars favoring low-mass stars as the most abundant. For each model along a track, theoretical asteroseismic mode frequencies were calculated up to the acoustic cut-off frequency with \texttt{ADIPLS}.
The individual frequencies are corrected for the effect of the asteroseismic surface effect \citep{Brown1984,ChristensenDalsgaard1984} following the cubic correction from \citet{BG14}.
The preferred value for each stellar parameter is adopted as the median of the marginalized posterior distribution, with uncertainties given by the 16th-84th percentile credibility interval.

\subsection{AMP~1.3} 

The AMP~1.3 results come from version 1.3 of the Asteroseismic Modeling Portal \citep[AMP;][]{Metcalfe2009}, which uses a parallel genetic algorithm \citep{Metcalfe2003} to optimize the stellar mass, composition, mixing-length, and age for a given set of observational constraints. This version of AMP relies on the Aarhus stellar evolution and adiabatic pulsation codes \citep{jcd08a, 2008Ap&SS.316..113C}, fitting frequency ratios in the same configuration as described by \cite{Creevey2017}. The input physics include the OPAL equation of state and opacities \citep{1996ApJ...464..943I, 2002ApJ...576.1064R} supplemented by the low temperature opacities of \cite{Ferguson2005}, the solar mixture of \cite{1998SSRv...85..161G}, the NACRE reaction rates \citep{angulo1999}, and the prescription of \cite{Michaud1993} to follow diffusion and settling of helium (but not heavy elements). Convection is described using the \cite{bohm1958} mixing-length prescription with no overshoot. The final parameter values and uncertainties are determined from a likelihood-weighted mean and standard deviation of all models sampled by the genetic algorithm in its search for an optimal reference model.

\subsection{\texttt{MESA}--\texttt{GYRE}}
The \texttt{MESA}--\texttt{GYRE} asteroseismic modeling results were obtained using stellar models computed with the stellar evolution code \texttt{MESA} \citep{2011ApJS..192....3P,2013ApJS..208....4P,2015ApJS..220...15P,Paxton2018,Paxton2019,Jermyn2023} (version r22.05.1). The model optimization relies on the differential evolution (DE) algorithm implemented in \texttt{yabox} \citep{Yabox}, which performs on-the-fly stellar modeling to identify parameters minimizing a cost function based on spectroscopic observables ($T_{\rm eff}$, [Fe/H], luminosity) and individual oscillation frequencies.

For each star, stellar tracks were computed while varying the initial mass ($0.8\le M_0 \le1.2$), helium mass fraction ($0.24\le Y_0 \le0.28$), metal-to-hydrogen ratio ($0.001\le Z_0/X_0\le0.06$), and mixing-length parameter ($1.6\le\alpha_{\rm mlt}\le2.0$). The models adopt the solar mixture of \citet{1998SSRv...85..161G} scaled to different metallicities, OPAL/Opacity Project opacities, and the default \texttt{MESA} equation-of-state (EOS) module blending several EOS sources—including
FreeEOS, OPAL/SCVH, HELM, and SKYE—\citep[see][]{Jermyn2023}. Elemental diffusion is included following \citet{Thoul1994}.

For each cost-function evaluation, a stellar track is evolved from the pre-main sequence to near hydrogen exhaustion ($X_{\rm center}=0.001$). Oscillation frequencies for radial, dipole, and quadrupole modes are then computed using \texttt{GYRE} v7.1 \citep{Townsend2013}. After applying the two-term surface correction of \citet{BG14}, the modeled frequencies and spectroscopic parameters ($T_\mathrm{eff}$, [Fe/H], and luminosity) are compared with observations to evaluate the total cost following the procedure of \citet{Lindsay2025}
\begin{equation}
\chi^2_{\rm total}=\chi^2_{\rm seismic}+\chi^2_{\rm spectroscopic}+\chi^2_{\rm low\,n},
\end{equation}
where $\chi^2_{\rm low\,n}$ is calculated using the two lowest-frequency radial and quadrupole modes.

The DE algorithm iteratively updates the model parameters over 40 iterations (840 cost-function evaluations). The model with the minimum $\chi^2_{\rm total}$ is adopted as the best-fit solution. Parameter uncertainties are estimated from likelihood-weighted standard deviations as described by \citet{Lindsay2025}.

\subsection{FICO}
The FICO procedure \citep[Forward and Inverse COmbination;][]{Betrisey2022,Betrisey2023_AMS_surf,Betrisey2024_AMS_quality, Betrisey2026} combines forward modeling and seismic inversions to mitigate surface effects and improve stellar parameter estimates. The method builds on the frameworks of \citet{Reese2012} and \citet{Buldgen2019f} and proceeds in three stages. We refer to \citet[][]{Betrisey2023_AMS_surf,Betrisey2026} for a comprehensive description of the methodology and a detailed description of the physics of the stellar models.

First, a reference model is obtained by fitting individual oscillation frequencies using the surface correction of \citet{BG14}. Although this model may contain biases due to imperfect surface-effect treatment, it provides a reliable estimate of the stellar mean density. In the second stage, this estimate is refined through a mean-density inversion using the nonlinear extension of the SOLA formalism \citep{Reese2012}, yielding a quasi model-independent constraint \citep[see][for a review]{Buldgen2022c}. The final stage performs a new modeling step using frequency separation ratios, which strongly suppress surface effects \citep{Roxburgh&Vorontsov2003,Oti2005}. While these ratios do not constrain the mean density directly, the inversion step compensates for this limitation.

Model optimization was carried out with the AIMS code \citep{Rendle2019} using the stellar model grid of \citet{Betrisey2026}. The models adopt the solar abundances of \citet{asplund2009}, OPAL opacities supplemented by low-temperature opacities from \citet{Ferguson2005}, the FreeEOS equation of state \citep{Irwin2012}, and nuclear reaction rates from \citet{Adelberger2011}. Microscopic diffusion follows \citet{Thoul1994} with screening coefficients from \citet{Paquette1986}. Convection is treated using mixing-length theory with $\alpha_{\rm mlt}=2.05$. Atmospheres are modeled using the $T(\tau)$ relation of \citet{Vernazza1981}.

The free parameters are stellar mass, age, and the initial helium and metal mass fractions ($Y_0$, $Z_0$). Non informative priors were adopted, except for age, for which we used a uniform prior between 0 and 13.8\,Gyr. The non-seismic constraints include $T_{\rm eff}$, surface metallicity, and luminosity. For the likelihood calculations, we modeled observational uncertainties as Gaussian noise and we attributed a weight of one to both non-seismic and seismic constraints. This means that the observational uncertainties were not altered (as would be, for example, with methods overweighting non-seismic constraints).

For solar analogs and at the precision levels required by missions such as PLATO \citep[Planetary Transits and Oscillations of Stars;][]{2014ExA....38..249R,2025ExA....59...26R}, both the direct frequency fitting and the full FICO procedure yield comparable results \citep[see e.g.][]{Betrisey2023_AMS_surf,Betrisey2026}. For consistency, we report two sets of results: the initial modeling step (FICO-1) and the complete procedure (FICO-3). The full procedure did not converge to a physical solution for EPIC~206371648, possibly due to unreliable observational constraints or missing physics in the model grid. This issue will be investigated in future work.

The frequency separation ratios could not be reliably fitted for KIC 3241581 because of limited mode coverage. Conversely, EPIC~212708252 exhibits a richer oscillation spectrum, leading to very tightly constrained posterior distributions when individual frequencies are fitted. As discussed by \citet{Rendle2019}, such high precision is expected for solar analogs and should not be interpreted as a numerical artifact. Nevertheless, we recommend adopting the FICO-3 results for further analysis, as they are less sensitive to this effect.

\subsection{Pitchfork}
We also derive stellar parameters using the \Pitchfork\ pipeline \citep{Scutt2026}. \Pitchfork\ is a multilayer perceptron neural network emulator of stellar evolution calculations designed to provide a fast and accurate alternative to grid interpolation.

The network was trained on the grid of stellar models presented by \citet{Lyttle2021}, consisting of 2.4 million models computed with \texttt{MESA} (version 12115). This grid uses the solar chemical mixture of $(Z/X)_\odot = 0.0181$ provided by \cite{asplund2009}, OPAL opacities and equation of state \citep{2002ApJ...576.1064R} supplemented by low-temperature opacities from \citet{Ferguson2005}, atomic diffusion of helium and heavy elements, and applies the \MESA predictive mixing scheme  \citep{Paxton2018,Paxton2019}.

We consider four model input parameters: mass \stellarmass, metallicity $Z_0$, helium abundance $Y_0$, and mixing length parameter \alphamlt, and then evolved forwards in steps of age \age. For each parameter set, the emulator predicts observable quantities including including \lum, \teff, and \surffeh, and 35 radial-mode frequencies ($6\le n\le40$) computed with \texttt{GYRE} \citep{Townsend2013}. The emulator achieves prediction precisions of 5.9\,K in $T_{\rm eff}$, $0.014\,L_\odot$ in luminosity, $6.5\times10^{-4}$\,dex in [Fe/H], and a mean frequency error of 0.02\%.

The trained network is used within a Bayesian inference framework to evaluate model likelihoods. Thanks to the millisecond-level evaluation time, we employ nested sampling with \UltraNest\ \citep{Buchner2021}. A Gaussian-process model following \citet{Li2023} accounts for correlated frequency residuals and allows the surface term parameters of the \citet{Kjeldsen2008} correction to be sampled simultaneously.

The computational scalability of \Pitchfork and the vectorized likelihood evaluation in \UltraNest allows us to go from stellar observable parameters to well sampled and fully marginalized posterior samples over 7 dimensions (\stellarmass, $Z_0$, $Y_0$, \alphamlt, \age, \KBa, \KBb) in minutes. To calculate the radius, we passed the posterior samples on the stellar fundamental properties back through \Pitchfork for the posterior predictive observables, and then converted the posterior predicted \teff and \lum to a posterior predicted radius using the Stefann-Boltzmann law. Our inferred value for a given stellar fundamental properties and the posterior predicted radius is the medians of the corresponding distribution, and the uncertainty is the 64~$\%$ confidence interval.

 \section{Discussion and conclusion}
 \label{sec:discussion}

\begin{figure}[!htb]
\begin{center}
    \includegraphics[width=0.5\textwidth, trim=0 0.6cm 0 0, clip]{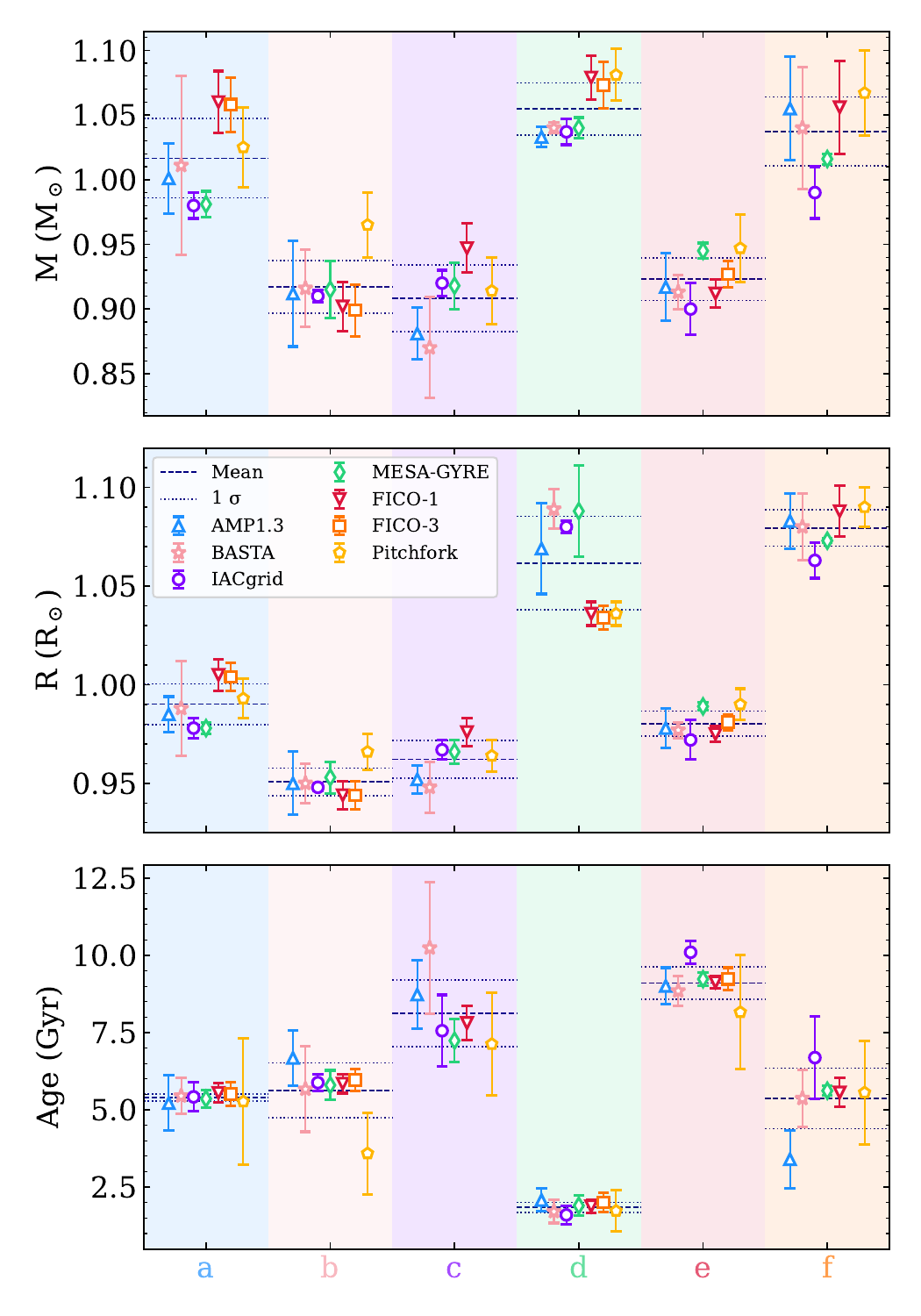}
\end{center}    
    \caption{Stellar parameters derived by the different stellar modeling methods for the six seismic solar analogs: mass (top panel), radius (middle panel), age (bottom panel). The letters for each star are given in Table~\ref{tab:models}. The dashed lines correspond to the mean value of the different models and the dotted lines correspond to the systematics.} 
    \label{fig:MRA}
\end{figure}

From the different stellar modeling methods described in Section~\ref{sec:modeling} and the available observables, we obtained the masses, radii, and ages of the six targets that are represented in Fig.~\ref{fig:MRA}. All the methods used as input the spectroscopic parameters ($\teff$, $\feh$, $\log g$), the luminosity, and the frequencies of the modes with an extended flag 0. The results from each method are given in Table~\ref{tab:models}.  For all the parameters, we find an overall good agreement between the different modeling codes. 

\begin{table}[!htb]
    \tiny
    \caption{Stellar fundamental parameters obtained for the six stars by the different modeling methods. }
\begin{center}
\begin{tabular}{cccc}
\hline
\hline
Method & $M$ (M$_\odot$)& $R$ (R$_\odot$) & Age (Gyr)\\ 
\hline
\\[-8pt]
\multicolumn{4}{c}{EPIC~206064678 (a)}\\
\hline
AMP1.3 &1.001\,$\pm$\,0.027	& 0.985\,$\pm$\,0.009	&5.22\,$\pm$\,0.89\\
\texttt{BASTA}	&1.011\,$\pm$\,0.069&	0.988\,$\pm$\,0.024&	5.45\,$\pm$\,0.58\\
IACgrid	&0.98\,$\pm$\,0.01	&0.978\,$\pm$\,0.005&	5.42\,$\pm$\,0.47\\
\texttt{MESA}-\texttt{GYRE}	&0.981\,$\pm$\,0.01	&0.978\,$\pm$\,0.003	&5.36\,$\pm$\,0.29\\
FICO-1 & 1.060\,$\pm$\,0.024 & 1.005\,$\pm$\,0.008 & 5.55\,$\pm$\,0.31\\
FICO-3 & 1.058\,$\pm$\,0.021 & 1.004\,$\pm$\,0.007 & 5.51\,$\pm$\,0.38\\
Pitchfork & 1.025\,$\pm$\,0.031 & 0.993\,$\pm$\,0.010 & 5.27\,$\pm$\,2.04\\
Mean & 1.016	&	0.990	&5.40\\
Systematics & 0.033	&0.011	&0.12\\
\hline
\\[-8pt]
\multicolumn{4}{c}{EPIC~206245055 (b)}\\
\hline
AMP1.3	&0.912	\,$\pm$\,0.041	&0.95\,$\pm$\,	0.016	&6.68\,$\pm$\,	0.9\\
\texttt{BASTA}	&0.916\,$\pm$\,	0.030	&0.95\,$\pm$\, 0.010	&5.67\,$\pm$\,	1.38\\
IACgrid	&0.91\,$\pm$\,0.005	&0.948\,$\pm$\,	0.002	&5.88\,$\pm$\,	0.26\\
\texttt{MESA}-\texttt{GYRE}	&0.915\,$\pm$\,	0.022	&0.953	\,$\pm$\,0.008	&5.80\,$\pm$\,	0.48\\
FICO-1 & 0.902\,$\pm$\,0.019 & 0.944\,$\pm$\,0.007 & 5.85\,$\pm$\,0.31\\
FICO-3 & 0.899\,$\pm$\,0.020 & 0.944\,$\pm$\,0.007 & 5.97\,$\pm$\,0.35\\
Pitchfork & 0.965\,$\pm$\,0.025 & 0.966\,$\pm$\,0.009 & 3.59\,$\pm$\,1.32\\
Mean	&0.917&		0.951&		5.63\\
Systematics	&0.022		&0.007		&0.96\\
\hline
\\[-8pt]
\multicolumn{4}{c}{EPIC~206371648 (c)}\\
\hline
AMP1.3&	0.881\,$\pm$\,0.02&	0.952\,$\pm$\,0.007&	8.73\,$\pm$\,1.11\\
\texttt{BASTA}&	0.870\,$\pm$\,0.039&	0.948\,$\pm$\,0.013&	10.24\,$\pm$\,2.12\\
IACgrid&	0.92\,$\pm$\,0.01&	0.967\,$\pm$\,0.005&	7.56\,$\pm$\,1.16\\
\texttt{MESA}-\texttt{GYRE}&	0.918\,$\pm$\,0.018&	0.966\,$\pm$\,0.006&	7.24\,$\pm$\,0.69\\
FICO-1 & 0.947\,$\pm$\,0.019 & 0.976\,$\pm$\,0.007 & 7.81\,$\pm$\,0.56 \\
FICO-3 & failed & failed & failed\\
Pitchfork & 0.914\,$\pm$\,0.026 & 0.964\,$\pm$\,0.008 & 7.13\,$\pm$\,1.66\\
Mean&	0.908&		0.962&		8.12	\\
Systematics&	0.028&		0.011&		1.18\\
\hline
\\[-8pt]
\multicolumn{4}{c}{EPIC~212624487 (d)}\\
\hline
AMP1.3 &1.033\,$\pm$\,0.008	&1.069	\,$\pm$\,0.023	&2.08	\,$\pm$\,0.37\\
\texttt{BASTA}	&1.04	\,$\pm$\,0.004	&1.089	\,$\pm$\,0.01	&1.71\,$\pm$\,0.37\\
IACgrid	&1.037	\,$\pm$\,0.01	&1.08	\,$\pm$\,0.003	&1.6	\,$\pm$\,0.3\\
\texttt{MESA}-\texttt{GYRE}	&1.04	\,$\pm$\,0.008	&1.088	\,$\pm$\,0.023	&1.90	\,$\pm$\,0.33\\
FICO-1 & 1.079\,$\pm$\,0.017 & 1.036\,$\pm$\,0.006 & 1.88\,$\pm$\,0.22 \\
FICO-3 & 1.073\,$\pm$\,0.018 & 1.034\,$\pm$\,0.006 & 2.01\,$\pm$\,0.31 \\
Pitchfork & 1.081\,$\pm$\,0.020 & 1.036\,$\pm$\,0.006 & 1.73\,$\pm$\,0.67\\
Mean	&1.043		&1.073		&1.84	\\
Systematics	&0.017		&0.018		&0.17	 \\
\hline
\\[-8pt]
\multicolumn{4}{c}{EPIC~212708252 (e)}\\
\hline
AMP1.3	&0.917\,$\pm$\,	0.026	&0.978\,$\pm$\,	0.01	&9.01\,$\pm$\,	0.59\\
\texttt{BASTA}	&0.913\,$\pm$\,	0.013	&0.977\,$\pm$\,	0.004	&8.85\,$\pm$\,	0.49\\
IACgrid	&0.90\,$\pm$\,	0.02	&0.972\,$\pm$\,	0.01	&10.1\,$\pm$\,	0.37\\
\texttt{MESA}-\texttt{GYRE}&	0.945\,$\pm$\,	0.006	&0.989\,$\pm$\,	0.002	&9.23\,$\pm$\,	0.22\\
FICO-1 & 0.912\,$\pm$\,0.011 & 0.975\,$\pm$\,0.004 & 9.13\,$\pm$\,0.19\\
FICO-3 & 0.927\,$\pm$\,0.010 & 0.981\,$\pm$\,0.004 & 9.24\,$\pm$\,0.36\\
Pitchfork & 0.947\,$\pm$\,0.026 & 0.990\,$\pm$\,0.008 & 8.16\,$\pm$\,1.84\\
Mean	&0.923		&0.948		&9.10	\\
Systematics	&0.018		&0.088	&	0.58	\\
\hline
\\[-8pt]
\multicolumn{4}{c}{KIC~3241581 (f)}\\
\hline
AMP1.3	&1.055\,$\pm$\,	0.04	&1.083\,$\pm$\,	0.014	&3.40\,$\pm$\,	0.93\\
\texttt{BASTA}	&1.040\,$\pm$\,	0.047	&1.080\,$\pm$\,	0.017	&5.37\,$\pm$\,	0.92\\
IACgrid&	0.99\,$\pm$\,	0.02	&1.063\,$\pm$\,	0.009	&6.69\,$\pm$\,	1.34\\
\texttt{MESA}-\texttt{GYRE}&	1.016\,$\pm$\,	0.004	&1.073\,$\pm$\,	0.001	&5.62\,$\pm$\,	0.16\\
FICO-1 & 1.056\,$\pm$\,0.036 & 1.088\,$\pm$\,0.013 & 5.56\,$\pm$\,0.47\\
FICO-3 & NA & NA & NA\\
Pitchfork & 1.067\,$\pm$\,0.033 & 1.090\,$\pm$\,0.010 & 5.56\,$\pm$\,1.68\\
Mean	&1.037		&1.080	&5.37	\\
Systematics	&0.029		&0.010		&1.07	\\
\hline
\end{tabular}
\end{center}
\label{tab:models}
\flushleft {\bf Notes.} ``Mean'' and ``Systematics'' correspond to the mean value and standard deviation from all methods respectively.  NA stands for Non~Applicable.
\end{table}

Table~\ref{tab:models} lists the mean values and systematics (estimated as the standard deviation) of the stellar parameters for each star. The inferred masses range from $\sim$0.91 to 1.04~M$_\odot$ and the radii from 0.95 to 1.08~R$_\odot$, confirming their close similarity to the Sun despite the use of different physics and optimization techniques. The derived ages span a wide range, from $\sim$1.8 to 9.1~Gyr. The ensemble comparison shown in Fig.~\ref{fig:MRA} indicates overall good agreement among the methods, even though the stellar evolution codes (\texttt{MESA}, \texttt{ASTEC}, and \texttt{GARSTEC}) adopt slightly different physics and surface-effect treatments, which may still contribute to some of the observed differences. The inferred stellar parameters ($M$, $R$, and age) are consistent within $1\sigma$ of the mean values when uncertainties are considered. Finally, none of the stars has a solar metallicity—except EPIC~212708252 (star e)—which differs from solar by several sigma.

\subsection{Star-by-star seismic modeling}
EPIC~206064678 (star a) is the star closest to the Sun in terms of its mean mass of 1.016\,$\mathrm{M}_\odot$ and radius of 0.99\,$\mathrm{R}_\odot$, taking into account the systematics between the different methods. However, it is slightly older than the Sun with a mean age of 5.40\,Gyr. Such a star is very rare and only a few solar twins with seismology have been studied. In comparison, the analysis of the known seismic solar twin, 18~Sco, by \citet{2018A&A...619A.172B} presented a bi-modality in the seismic age modeling of 4.67 and 6.95\,Gyr, implying a large age uncertainty. We note that the inferred masses and radii for this star (a) cluster in three groups: three models are around the mean, two are below 1\,$\sigma$ and the two solutions of FICO are above 1\,$\sigma$. They correspond to methods that use three different stellar evolution codes, which could explain the differences.

All models of EPIC~206245055 (star b) have a very good agreement in all the parameters shown in Fig.~\ref{fig:MRA} except for \texttt{Pitchfork} but the parameters inferred with this model are still at the level of 1\,$\sigma$ agreement.

As shown in the \'echelle diagram of EPIC~206371648 (star c) (Figure~\ref{fig:echelle_206371648}), only five $\ell$\,=\,2 modes were fitted, with all of them having very large error bars. The ridge of these quadrupolar modes is not straight (not following the $\ell$\,=\,0 ridge). This means that the small separation $\delta\nu_{0,2}$, which is sensitive to the properties of the core, is not well constrained. This has a direct impact on the mass and age retrieved by the best-fit model found by each method, yielding a larger discrepancy in these stellar parameters.

The inferred mass and radius of EPIC~212624487 (star d) split into two solutions, with one favoring slightly higher mass and smaller radius and the other favoring lower mass and larger radius. Interestingly, the ages converge towards a very similar value. Together with EPIC~206064678 (star a), this star shows the smallest scatter in the age determination.

EPIC~2127088252 (star e) has a very good agreement in the stellar parameters derived by the different methods. It is the oldest star of the sample with a mean age of 9.1\,Gyr.

The \emph{Kepler} target KIC~3241581 (star f) shows a large dispersion in $M$ and age, with correspondingly large uncertainties returned by the models. This likely reflects the limited seismic constraints available for this star: only eight modes are detected, with relatively large frequency uncertainties, and only a single $\ell=2$ mode. As a result, the stellar core properties are only weakly constrained. The derived stellar parameters should therefore be treated with increased caution. Indeed, when adopting the full set of modes reported by \texttt{apollinaire}, the modeling converges toward a significantly different solution, with a higher mass ($\sim1.125\,\mathrm{M}_\odot$), a larger radius ($\sim1.110\,\mathrm{R}_\odot$), and a younger age ($\sim5.07$\,Gyr). Interestingly, while the inferred mass and radius change noticeably, the age varies much less. This suggests that the age may not be the least well-constrained parameter for this target, in a way similar to what is found for star~d.

\subsection{Surface magnetic activity}

\begin{figure}[!htb]
\begin{center}
    \includegraphics[width=\columnwidth]{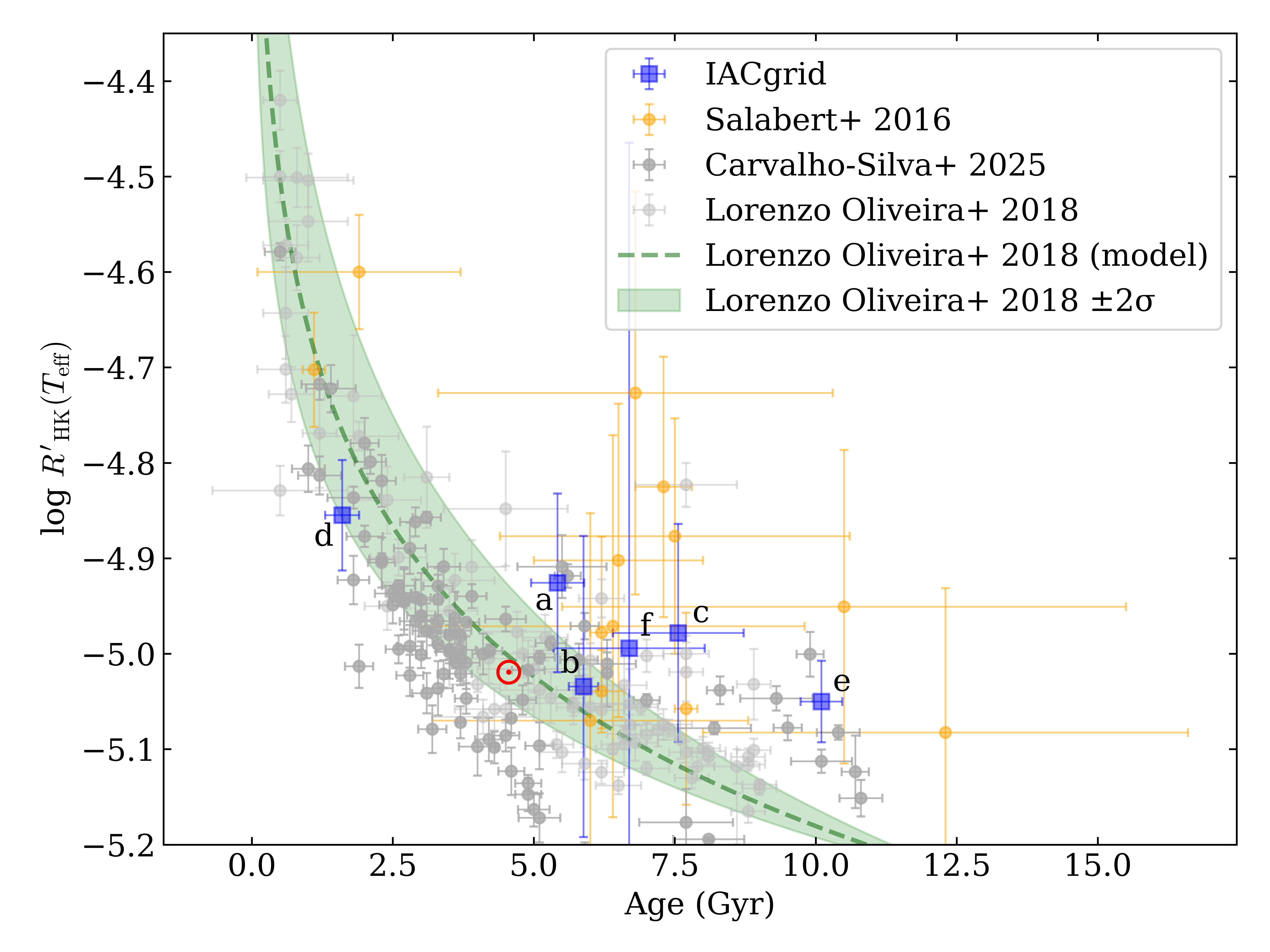}
\end{center}    
    \caption{$\log R'_\mathrm{HK}(T_\mathrm{eff})$ as a function of age for solar twins and solar analogs. Blue squares represent the six new seismic solar analogs listed in Table~\ref{tab:Seismic}. Orange, dark-gray, and light-gray circles correspond to stellar samples from the literature, as indicated in the legend. The dashed line shows the empirical relation from \cite{2018A&A...619A..73L}. The green shaded region represents the $2\sigma$ uncertainty envelope around this relation. The Sun is represented by its red symbol.} 
    \label{fig:ensemble_activity}
\end{figure}

This sample of stars is particularly interesting because, despite stellar parameters within about 10\% of solar values and spanning a broader range of metallicities, all stars exhibit exceptionally low magnetic activity \citep[see for example,][]{2016A&A...596A..31S,2018A&A...619A..73L,2025ApJ...983L..31C}. While previous studies on the magnetic activity of solar-like stars shown that they had enhanced magnetic activity compared to the Sun, they mostly focused on young fast rotators \citep[e.g.][]{2018A&A...616A.108B}. Recent studies have shown that the Sun exhibits a level of magnetic activity comparable to that of other solar-like stars \citep[e.g.][]{2021ApJ...908L..21R,2023ApJ...952..131M,2023A&A...672A..56S,2025ApJ...982..114M}. 
The magnetic activity indexes (S-index, $\log R'_\mathrm{HK} (B-V)$, and $\log R'_\mathrm{IRT}$) computed in Section~\ref{sec:spectroscopic_analysis} are in a similar range of the solar value between minimum and maximum activity.  This is not surprising as it is known that the seismic sample is biased towards stars with lower magnetic activity. Indeed, strong level of magnetic activity can be responsible for lower mode amplitude \citep[e.g.][]{2010Sci...329.1032G, 2011ApJ...732L...5C, 2019FrASS...6...46M,2025A&A...700A..25B}. 
To put our sample in context of the solar twins and solar analogs studied by \citet{2018A&A...619A..73L} and \citet{2025ApJ...983L..31C}, we converted $\log R'_{\rm HK}(B-V)$ to the $T_\mathrm{eff}$ scale, $\log R'_\mathrm{HK}(T_\mathrm{eff})$, following the prescription of \citet{2018A&A...619A..73L} and using our measured $S$-index values. Figure~\ref{fig:ensemble_activity} shows the evolution of this activity index as a function of age for different samples of stars. Our youngest star, (d), is well within the empirical relation derived by \citep{2018A&A...619A..73L} for solar twins. The remaining stars that are older than the Sun are in general outside the 2\,$\sigma$ envelope. Indeed, since our sample is not made up of solar twins but of solar analogs, they are populating the region covered by the sample of 
 \citet{2025ApJ...983L..31C}, where differences could arise from the metallicities of our targets. These stars are consistent with the existence of a plateau of magnetic activity for old stars. Compared to the solar analogs from \citet{2016A&A...596A..31S}, the age uncertainties are much smaller, thanks to the individual mode seismic modeling in contrast to the global seismic parameter modeling \citep[see also,][]{Grossmann26}.

It is important to note that our inferred stellar parameters have internal uncertainties larger than those of previous works on solar analogs and twins based solely on spectroscopic analysis \citep[e.g.][]{2018A&A...619A..73L,2025ApJ...983L..31C}. This can be partially explained by the differential spectroscopic analysis yielding very small error bars in the temperature smaller than 10~K, compared to our standard analysis providing typical values around 50~K.

\subsection{Evolution of Lithium} 

\begin{figure}[!t]
    \centering
    \includegraphics[width=\linewidth]{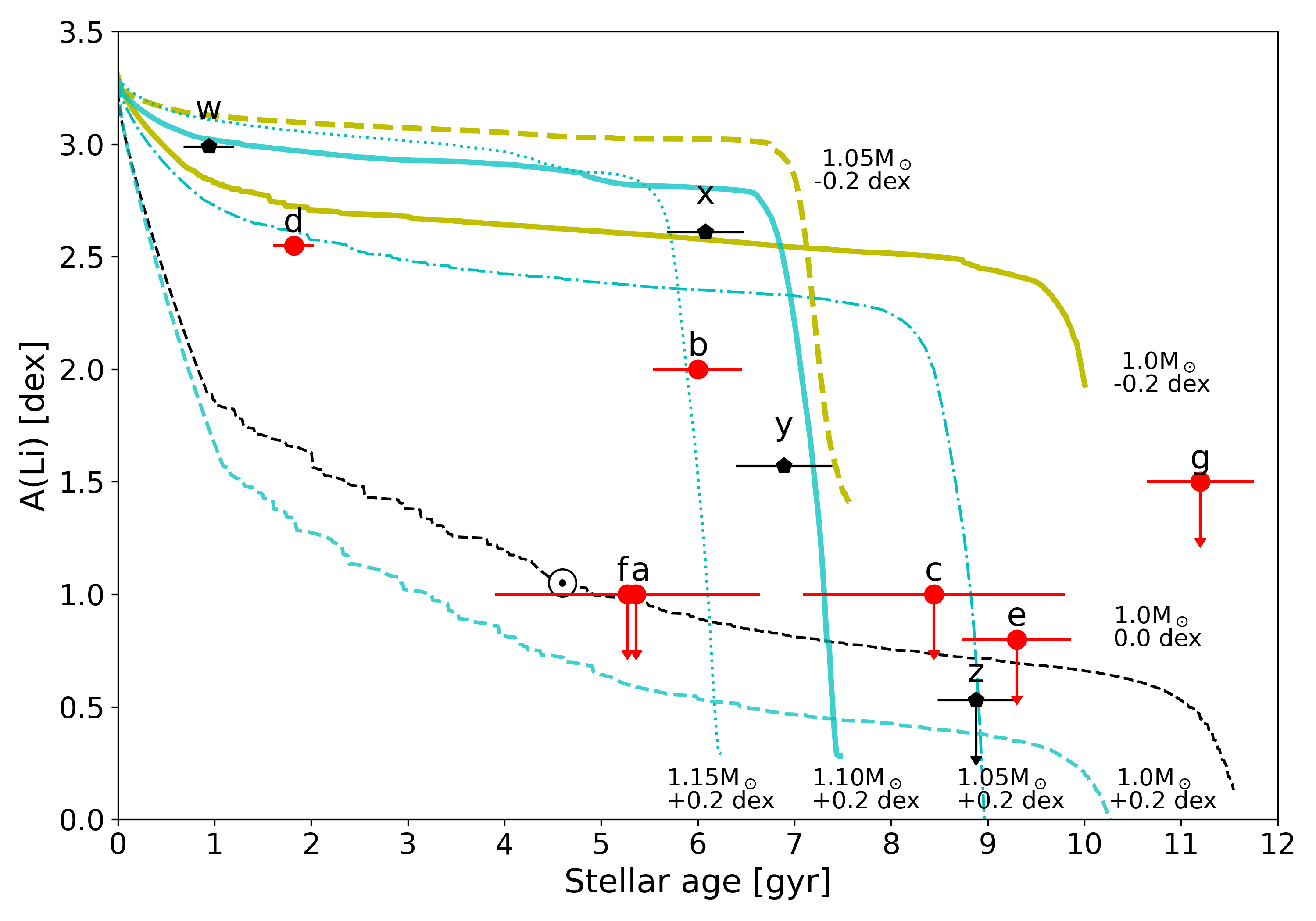}
    \caption{ Lithium abundance of single stars with age computed by detailed
modeling with the amp code. 
The stars from this work are shown as red dots. Additional stars:
(g) KIC\,9693187  from \cite{2026A&A...706L..19B} and 
(w) KIC\,10644253, 
(x) KIC\,6116048,
(y) KIC\,7680114, and 
(z) KIC\,3656476  from \cite{Beck2017} are depicted as black pentagons. 
Vertical, down-oriented errors indicate that the reported A(Li) is an upper limit (for stars a,c,e,f,g, and z). 
TGECevolutionary tracks from \cite{Beck2017} are shown in cyan, and yellow represents the theoretical evolution of A(Li) for models of the indicated mass with
[Fe/H] = +0.2, and -0.2 dex, respectively. The black dashed evolutionary track depicts the evolution of Li calculated by do \cite{doNascimento2009}, and the solar marker depicts the measured A(Li$_\odot$) and age of the
Sun.}
    \label{fig:LithiumAge}
\end{figure}

The availability of precise asteroseismic ages for this sample provides new constraints on the long-term evolution of lithium in solar analogs. As shown in Fig.~\ref{fig:LithiumAge}, all six stars display lithium abundances well below the meteoritic value, with several targets providing only upper limits, despite spanning ages of~$\sim$1.8–9.1~Gyr and a range of metallicities. This is consistent with previous seismic studies showing that substantial lithium depletion occurs already on the main sequence and continues during solar-type evolution, with a dispersion that cannot be explained by age alone \citep{Beck2017}.
The binary periods exclude strong tidal angular momentum transport \citep{Zahn1977, Zahn1989}, allowing the stars to be treated as effectively single. In this context, lithium depletion is likely driven by non-standard internal transport processes linked to stellar rotation and angular momentum evolution. Recent theoretical work supports this interpretation, showing that lithium evolution in solar-type stars is highly sensitive to the efficiency and depth of internal mixing and requires physics beyond standard solar-calibrated models \citep{2005Sci...309.2189C,doNascimento2009,Buldgen2025}.
The very low magnetic activity levels observed in all targets indicate that surface activity alone cannot explain the depletion. In particular, EPIC~206064678—the star most similar to the Sun in mass and radius—shows a lithium abundance comparable to or lower than the solar value despite being slightly older and more metal rich, supporting the conclusion that the Sun is not anomalously lithium-poor in a seismic evolutionary context. A more detailed analysis will be presented for a larger sample in future work.

In summary, using the set of stars analyzed in this work, we have extended the parameter space of seismic solar analogs to encompass a wider range of metallicities and low magnetic activity levels. Broadening this parameter space is essential for improving our understanding of the Sun’s evolution and placing it in a broader context, especially in anticipation of the PLATO mission, scheduled for launch in early 2027, which will seismically characterize thousands of solar-type stars, many of which are expected to be solar analogs spanning a wide range of parameters.

\section{Data availability}
A single table containing the individual frequencies for all the stars given in Appendix~\ref{App:extraction} is available in electronic form at the CDS via anonymous ftp to \url{cdsarc.u-strasbg.fr} (130.79.128.5) or via \url{http://cdsweb.u-strasbg.fr/cgi-bin/qcat?J/A+A/}.

\begin{acknowledgements}
This paper includes data collected by the \emph{Kepler}/K2 missions. Funding for the \emph{Kepler} mission is provided by the NASA Science Mission directorate. Some of the data presented in this paper were obtained from the Mikulski Archive for Space Telescopes (MAST). STScI is operated by the Association of Universities for Research in Astronomy, Inc., under NASA contract NAS5-26555. This paper also contains observations obtained with the HERMES spectrograph, which is supported by the Research Foundation – Flanders (FWO), Belgium, the Research Council of KU Leuven, Belgium, the Fonds National de la Recherche Scientifique (F.R.S.-FNRS), Belgium, the Royal Observatory of Belgium, the Observatoire de Genève, Switzerland and the Thüringer Landessternwarte Tautenburg, Germany.
R.A.G., E.P., B.L., D.B.P., A.R.G.S. and V.D. acknowledge the support from the GOLF and PLATO Centre National D'{\'{E}}tudes Spatiales grants. 
S.M. acknowledges support by the Spanish Ministry of Science and Innovation through AEI under the Severo Ochoa Centres of Excellence Programme 2020--2023 (CEX2019-000920-S). 
S.N.B acknowledges support from PLATO ASI-INAF agreement n.~2015-019-R.1-2018. P.G.B. acknowledges support by the Spanish Ministry of Science and Innovation with the \textit{Ram{\'o}n\,y\,Cajal} fellowship number RYC-2021-033137-I and the number MRR4032204. P.G.B., D.H.G., T.M., C.A.P., D.G.R. and R.A.G. acknowledge support from the Spanish Ministry of Science and Innovation with the grant no. PID2023-146453NB-100 (\textit{PLAtoSOnG}).
S.M., D.G.R., D.H.G., and R.A.G. acknowledge support from the Spanish Ministry of Science and Innovation with the grant no. PID2023-149439NB-C41. D.G.R. acknowledges support from the Spanish Ministry of Science and Innovation (MICINN) with the Juan de la Cierva fellowship program under contract JDC2022-049054-I.
The research of J.M. was supported by the Czech Science Foundation (GACR) project no. 24-10608O.
J.B. acknowledges funding from the SNF Postdoc.Mobility grant no. P500PT{\_}222217. GB acknowledges fundings from the Fonds National de la Recherche Scientifique (FNRS) as a postdoctoral researcher.
AE received the support of a fellowship from "La Caixa" Foundation (ID 100010434) with fellowship code LCF/BQ/PI23/11970031. 
M.N.L. acknowledges support from the ESA PRODEX programme (PEA 4000142995). T.S.M. acknowledges support from NASA grant 80NSSC25K7563. Computational time at the Texas Advanced Computing Center was provided through ACCESS allocation TG-AST090107.
This research was supported by the International Space Science Institute (ISSI) in Bern, through the ISSI International Team project 24-629 (\textit{"Multi-scale variability in solar and stellar magnetic cycles}")
O.J.S. acknowledges the support of the Science and Technology Facilities Council (STFC).
The authors wish to acknowledge the contribution of the IAC  High-Performance Computing support team and hardware facilities to the results of this research.

\\

\textit{Software:}   
\texttt{ADIPLS} \citep{2008Ap&SS.316..113C},
AIMS code \citep{Rendle2019},
AMP \citep{Metcalfe2009},
\texttt{apollinaire} \citep{2022A&A...663A.118B}, 
AstroPy \citep{astropy:2013,astropy:2018}, 
\texttt{BASTA} \citep[][]{aguirre22},
ChatGPT \citep{openai2023chatgpt}, 
emcee \citep{2013PASP..125..306F}, 
FICO \citep{Betrisey2022,Betrisey2023_AMS_surf,Betrisey2024_AMS_quality,Betrisey2026},
\texttt{garstec}; \citep{weiss2008},
\texttt{GYRE} \citep{Townsend2013},
Interactive Data Language (IDL\footnote{\url{https://www.nv5geospatialsoftware.com/docs/home.html}}), 
\texttt{iechelle}\footnote{\url{https://gitlab.com/dinilbose/iechelle}} (Palakkatharappil, García, et al., in preparation),
Matplotlib \citep{matplotlib}, 
\texttt{MESA} \citep{2011ApJS..192....3P,2013ApJS..208....4P,2015ApJS..220...15P},
NumPy \citep{numpy}, 
pandas \citep{mckinney-proc-scipy-2010,reback2020pandas}, 
\texttt{PBjam} \citep{2021AJ....161...62N, Nielsen_2025},
\Pitchfork\ \citep{Scutt2026},
SciPy \citep{scipy}, 
\UltraNest\ \citep{Buchner2021},
\texttt{yabox} \citep{Yabox}.

\end{acknowledgements}

%
%
\bibliographystyle{aa.bst}

\bibliography{BIBLIO.bib} 

\begin{appendix} 
\label{App:peak-bag_results}

\section{Summary of the p-mode frequencies}
\label{App:extraction}

\subsection{KIC 3241581}

\begin{table}[H]
\centering
\begin{tabular}{ccccc}
\hline
$n$ & $\ell$ & $\nu_{\texttt{apollinaire}}$ & $\nu_{\mathrm{PBJAM}}$ & extended \\
\hline
16 & 0 & 2149.20 $\pm$ 1.40 & & 1 \\
17 & 0 & 2268.53 $\pm$ 1.33 & & 1 \\
18 & 0 & 2388.95 $\pm$ 1.45 & & 1 \\
19 & 0 & 2509.39 $\pm$ 0.58 & & 0 \\
20 & 0 & 2632.97 $\pm$ 0.92 & & 0 \\
21 & 0 & 2753.93 $\pm$ 0.65 & 2754.29 $\pm$ 1.64 & 0 \\
23 & 0 & 3001.12 $\pm$ 1.30 & 3000.66 $\pm$ 2.12 & 0 \\
24 & 0 & 3125.08 $\pm$ 1.31 & 3123.76 $\pm$ 2.83 & 0 \\
16 & 1 & 2202.00 $\pm$ 0.98 & & 1 \\
19 & 1 & 2567.08 $\pm$ 0.67 & & 1 \\
20 & 1 & 2689.77 $\pm$ 0.13 & 2689.80 $\pm$ 1.12 & 0 \\
21 & 1 & 2812.75 $\pm$ 0.44 & 2812.69 $\pm$ 0.77 & 0 \\
22 & 1 & 2935.70 $\pm$ 0.40 & 2935.83 $\pm$ 0.46 & 0 \\
24 & 1 & 3181.60 $\pm$ 1.03 & 3181.65 $\pm$ 0.75 & 0 \\
18 & 2 & 2500.56 $\pm$ 1.09 & & 1 \\
19 & 2 & 2622.33 $\pm$ 0.71 & 2623.74 $\pm$ 3.15 & 0 \\
\hline
\end{tabular}
\caption{Frequencies and associated incertitude for each detected modes $(n, l)$, and depending on the peakbagging code, for KIC 3241581. The extended column takes a value of 0 when the mode frequencies obtained from both fitting methods are consistent within 1$\sigma$, and 1 otherwise.}
\label{tab:freq_3241581}
\end{table}
\vspace{-30pt}
\begin{figure}[H]
    \centering
    \includegraphics[width=0.9\linewidth]{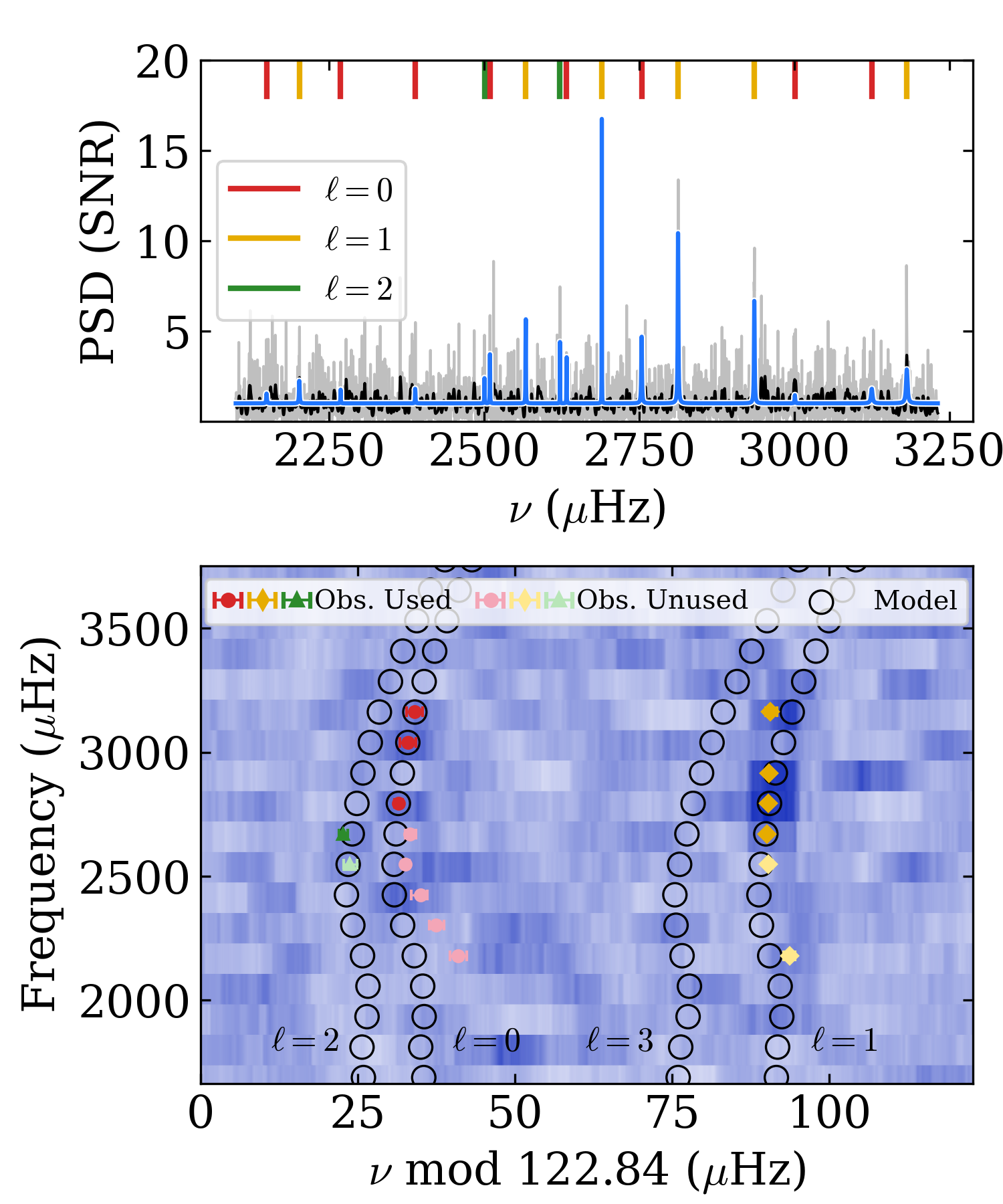}
    \caption{Left panel: Power spectral density (PSD) of KIC 3241581 in the region of the oscillation power excess. The full excess is shown in SNR units, with the original PSD in black and its smoothed version (using a 10-point window) in grey. The blue curve represents the combined model based on the  frequencies extracted by \texttt{apollinaire}, which are marked by vertical dashes: red for $\ell=0$, yellow for $\ell=1$, and green for $\ell=2$. Right panel: Corresponding échelle diagram. Observed mode frequencies are shown as filled symbols where darker ones indicate the frequencies included in the model, while lighter ones were excluded. Color coding is consistent with the left panel: red circles, yellow diamonds, and green triangles represent $\ell=0$, $\ell=1$, and $\ell=2$, respectively. Model frequencies are plotted as open black symbols with consistent markers. This figure was produced using the Python module \texttt{pareidolia} (available at \url{https://gitlab.com/evapanetier/pareidolia.git}), which is designed for seismic data analysis. }
    \label{fig:echelle_3241581}
\end{figure}

\subsection{EPIC 206064678}

\begin{table}[H]
\centering
\begin{tabular}{ccccc}
\hline
$n$ & $\ell$ & $\nu_{\texttt{apollinaire}}$ & $\nu_{\mathrm{PBJAM}}$ & extended \\
\hline
17 & 0 & 2558.81  $\pm$ 2.67 & 2557.92  $\pm$ 3.07 & 0 \\
18 & 0 & 2696.27  $\pm$ 0.41 & 2696.01  $\pm$ 1.04 & 0 \\
19 & 0 & 2834.67  $\pm$ 1.10 & 2834.41  $\pm$ 1.09 & 0 \\
20 & 0 & 2971.46  $\pm$ 0.30 & 2971.43  $\pm$ 0.35 & 0 \\
21 & 0 & 3109.32  $\pm$ 0.19 & 3109.38  $\pm$ 0.29 & 0 \\
22 & 0 & 3247.13  $\pm$ 0.22 & 3247.11  $\pm$ 0.23 & 0 \\
23 & 0 & 3385.22  $\pm$ 0.79 & 3385.95  $\pm$ 3.06 & 0 \\
24 & 0 & 3523.14  $\pm$ 0.47 & 3523.18  $\pm$ 0.86 & 0 \\
25 & 0 & 3660.10  $\pm$ 2.15 & & 1 \\
16 & 1 & 2486.84  $\pm$ 1.56 & 2487.57  $\pm$ 3.79 & 0 \\
18 & 1 & 2761.57  $\pm$ 0.26 & 2761.53  $\pm$ 0.31 & 0 \\
19 & 1 & 2899.20  $\pm$ 0.12 & 2899.29  $\pm$ 0.14 & 0 \\
20 & 1 & 3037.52  $\pm$ 0.20 & 3037.45  $\pm$ 0.20 & 0 \\
21 & 1 & 3175.17  $\pm$ 0.19 & 3175.15  $\pm$ 0.21 & 0 \\
22 & 1 & 3313.63  $\pm$ 0.20 & 3313.56  $\pm$ 0.23 & 0 \\
23 & 1 & 3452.05  $\pm$ 0.44 & 3452.09  $\pm$ 0.41 & 0 \\
24 & 1 & 3590.33  $\pm$ 0.30 & 3590.23  $\pm$ 1.56 & 0 \\
18 & 2 & 2824.13  $\pm$ 0.28 & 2824.22  $\pm$ 1.15 & 0 \\
19 & 2 & 2962.63  $\pm$ 5.05 & & 1 \\
20 & 2 & 3099.75  $\pm$ 0.46 & 3099.79  $\pm$ 1.04 & 0 \\
21 & 2 & 3238.99  $\pm$ 1.59 & 3239.88  $\pm$ 2.91 & 0 \\
22 & 2 & 3376.87  $\pm$ 0.95 & 3377.05  $\pm$ 0.55 & 0 \\
23 & 2 & 3517.93  $\pm$ 1.88 & & 1 \\
\hline
\end{tabular}
\caption{Same as in Table~\ref{tab:freq_3241581} shown here for EPIC 206064678}
\label{tab:freq_206064678}
\end{table}

\begin{figure}[H]
    \centering
    \includegraphics[width=0.9\linewidth]{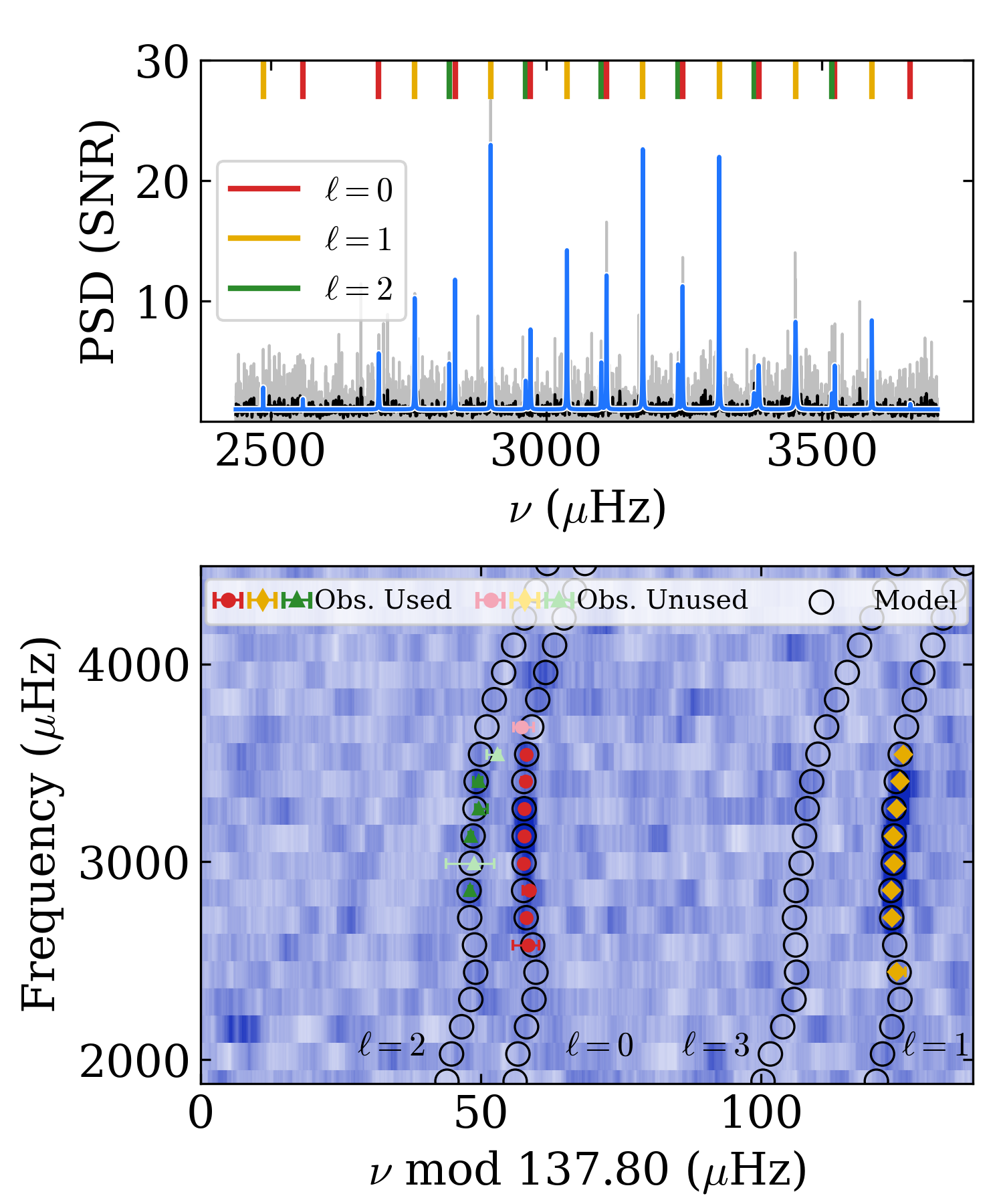}
    \caption{Same legend as in Figure~\ref{fig:echelle_3241581}; shown here for EPIC 206064678.}
    \label{fig:echelle_206064678}
\end{figure}

\subsection{EPIC 206245055}

\begin{table}[H]
\centering
\begin{tabular}{ccccc}
\hline
$n$ & $\ell$ & $\nu_{\texttt{apollinaire}}$ & $\nu_{\mathrm{PBJAM}}$ & extended \\
\hline
17 & 0 & 2566.18 $\pm$ 1.47 & 2566.19 $\pm$ 2.47 & 0 \\
18 & 0 & 2703.63 $\pm$ 0.94 & 2703.46 $\pm$ 1.95 & 0 \\
19 & 0 & 2843.67 $\pm$ 0.33 & 2843.67 $\pm$ 0.35 & 0 \\
20 & 0 & 2981.75 $\pm$ 0.31 & 2981.82 $\pm$ 0.28 & 0 \\
21 & 0 & 3120.51 $\pm$ 0.23 & 3120.51 $\pm$ 0.22 & 0 \\
22 & 0 & 3259.84 $\pm$ 0.62 & 3259.75 $\pm$ 1.16 & 0 \\
23 & 0 & 3397.15 $\pm$ 1.93 & 3397.44 $\pm$ 2.72 & 0 \\
24 & 0 & 3538.58 $\pm$ 0.50 & 3538.64 $\pm$ 1.20 & 0 \\
25 & 0 & 3679.34 $\pm$ 3.31 & 3679.48 $\pm$ 2.89 & 0 \\
26 & 0 & 3820.04 $\pm$ 3.77 & 3818.68 $\pm$ 3.47 & 0 \\
27 & 0 & 3962.54 $\pm$ 3.00 & & 1 \\
15 & 1 & 2356.51 $\pm$ 0.88 & 2356.01 $\pm$ 2.22 & 0 \\
16 & 1 & 2494.21 $\pm$ 1.33 & 2494.16 $\pm$ 1.12 & 0 \\
17 & 1 & 2632.01 $\pm$ 0.70 & 2632.10 $\pm$ 0.60 & 0 \\
18 & 1 & 2769.82 $\pm$ 0.70 & 2770.03 $\pm$ 0.61 & 0 \\
19 & 1 & 2909.16 $\pm$ 0.35 & 2908.80 $\pm$ 0.41 & 0 \\
20 & 1 & 3048.71 $\pm$ 0.40 & 3048.83 $\pm$ 0.43 & 0 \\
21 & 1 & 3187.86 $\pm$ 0.45 & 3187.71 $\pm$ 0.41 & 0 \\
22 & 1 & 3326.23 $\pm$ 0.23 & 3326.14 $\pm$ 0.24 & 0 \\
23 & 1 & 3466.65 $\pm$ 0.63 & 3466.60 $\pm$ 0.66 & 0 \\
24 & 1 & 3607.64 $\pm$ 0.50 & 3607.58 $\pm$ 0.58 & 0 \\
25 & 1 & 3747.31 $\pm$ 2.81 & 3746.56 $\pm$ 3.23 & 0 \\
16 & 2 & 2557.52 $\pm$ 2.01 & 2558.04 $\pm$ 3.80 & 0 \\
17 & 2 & 2695.78 $\pm$ 1.00 & 2696.35 $\pm$ 1.20 & 0 \\
18 & 2 & 2834.21 $\pm$ 1.71 & 2835.09 $\pm$ 2.97 & 0 \\
19 & 2 & 2973.76 $\pm$ 1.16 & 2973.52 $\pm$ 1.23 & 0 \\
20 & 2 & 3111.82 $\pm$ 1.00 & 3111.89 $\pm$ 2.32 & 0 \\
21 & 2 & 3251.77 $\pm$ 2.55 & 3251.64 $\pm$ 2.96 & 0 \\
22 & 2 & 3390.19 $\pm$ 2.13 & 3391.18 $\pm$ 2.70 & 0 \\
23 & 2 & 3531.99 $\pm$ 0.52 & 3532.18 $\pm$ 2.61 & 0 \\
24 & 2 & 3671.25 $\pm$ 2.73 & 3671.33 $\pm$ 3.22 & 0 \\
\hline
\end{tabular}
\caption{Same as in Table~\ref{tab:freq_3241581} shown here for EPIC 206245055.}
\label{tab:freq_206245055}
\end{table}

\begin{figure}[H]
    \centering
    \includegraphics[width=0.9\linewidth]{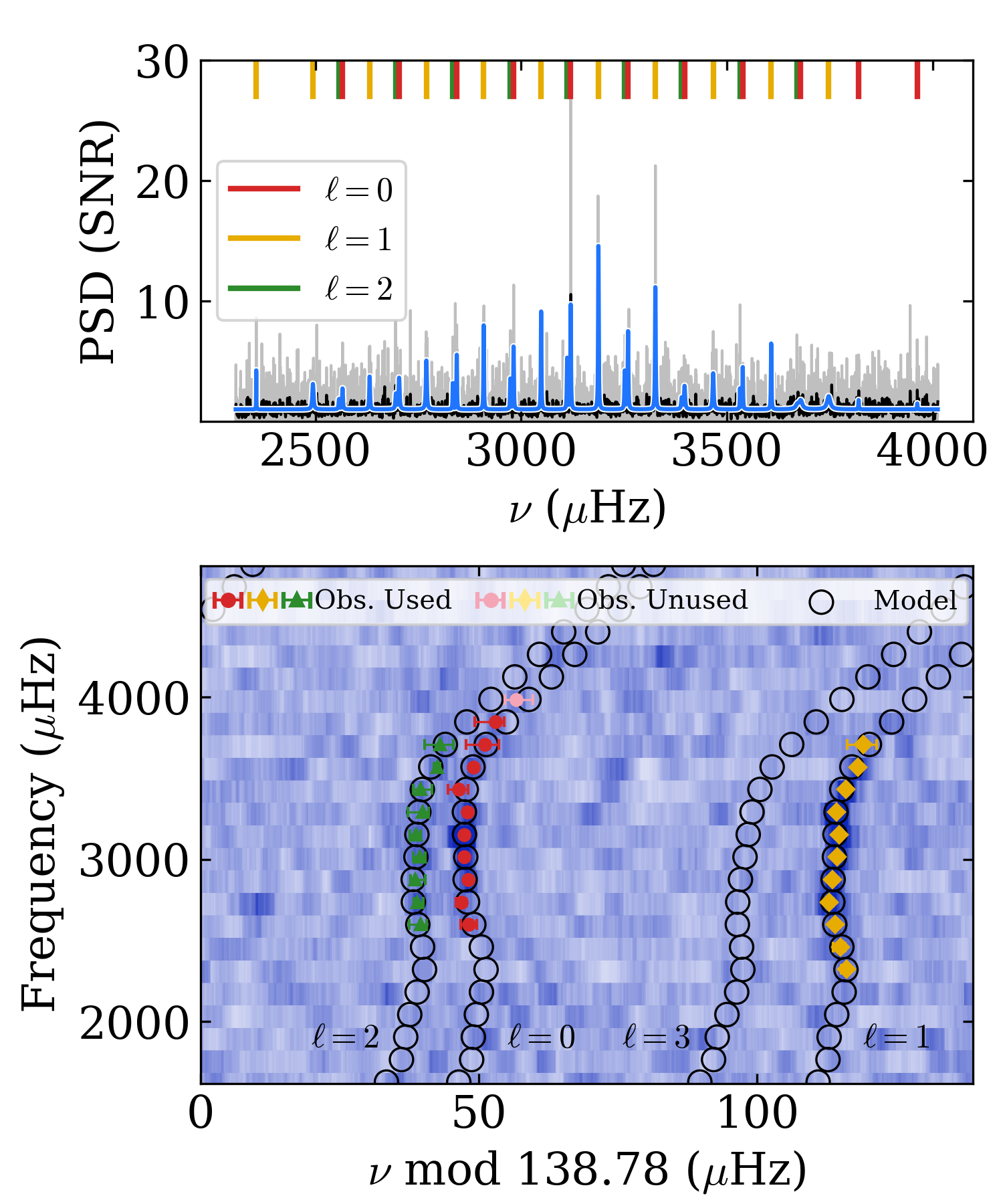}
    \caption{Same legend as in Figure~\ref{fig:echelle_3241581}; shown here for EPIC~206245055.}
    \label{fig:echelle_206245055}
\end{figure}

\subsection{EPIC 206371648}

\begin{table}[H]
\centering
\begin{tabular}{ccccc}
\hline
$n$ & $\ell$ & $\nu_{\texttt{apollinaire}}$ & $\nu_{\mathrm{PBJAM}}$ & extended \\
\hline
14 & 0 & 2114.89 $\pm$ 1.51 & & 1 \\
17 & 0 & 2519.22 $\pm$ 0.18 & 2519.21 $\pm$ 2.84 & 0 \\
18 & 0 & 2653.96 $\pm$ 1.22 & 2654.05 $\pm$ 2.78 & 0 \\
19 & 0 & 2790.38 $\pm$ 0.48 & 2789.89 $\pm$ 1.98 & 0 \\
20 & 0 & 2926.84 $\pm$ 0.61 & 2926.56 $\pm$ 1.57 & 0 \\
21 & 0 & 3062.67 $\pm$ 0.26 & 3062.61 $\pm$ 0.36 & 0 \\
22 & 0 & 3198.52 $\pm$ 2.58 & 3198.44 $\pm$ 0.94 & 0 \\
23 & 0 & 3335.62 $\pm$ 2.27 & 3335.23 $\pm$ 2.70 & 0 \\
24 & 0 & 3471.86 $\pm$ 0.24 & 3471.89 $\pm$ 0.23 & 0 \\
25 & 0 & 3607.31 $\pm$ 3.91 & & 1 \\
17 & 1 & 2583.32 $\pm$ 0.45 & 2583.41 $\pm$ 1.11 & 0 \\
18 & 1 & 2719.41 $\pm$ 0.40 & 2719.58 $\pm$ 0.39 & 0 \\
19 & 1 & 2855.50 $\pm$ 0.21 & 2855.39 $\pm$ 0.26 & 0 \\
20 & 1 & 2991.95 $\pm$ 0.23 & 2991.99 $\pm$ 0.36 & 0 \\
21 & 1 & 3128.15 $\pm$ 0.18 & 3128.24 $\pm$ 0.25 & 0 \\
22 & 1 & 3264.52 $\pm$ 0.58 & 3264.45 $\pm$ 0.58 & 0 \\
23 & 1 & 3401.44 $\pm$ 3.61 & 3400.74 $\pm$ 2.78 & 0 \\
24 & 1 & 3541.25 $\pm$ 1.81 & 3539.21 $\pm$ 3.61 & 0 \\
25 & 1 & 3678.40 $\pm$ 2.59 & 3676.33 $\pm$ 3.87 & 0 \\
17 & 2 & 2646.82 $\pm$ 2.92 & 2646.91 $\pm$ 3.81 & 0 \\
18 & 2 & 2786.58 $\pm$ 1.70 & 2785.39 $\pm$ 3.29 & 0 \\
20 & 2 & 3055.60 $\pm$ 2.01 & 3055.13 $\pm$ 2.18 & 0 \\
22 & 2 & 3325.78 $\pm$ 2.29 & 3326.62 $\pm$ 1.26 & 0 \\
23 & 2 & 3462.57 $\pm$ 0.92 & 3462.57 $\pm$ 3.07 & 0 \\
\hline
\end{tabular}
\caption{Same as in Table~\ref{tab:freq_3241581} shown here for EPIC 206371648.}
\label{tab:freq_206371648}
\end{table}

\begin{figure}[H]
    \centering
    \includegraphics[width=0.9\linewidth]{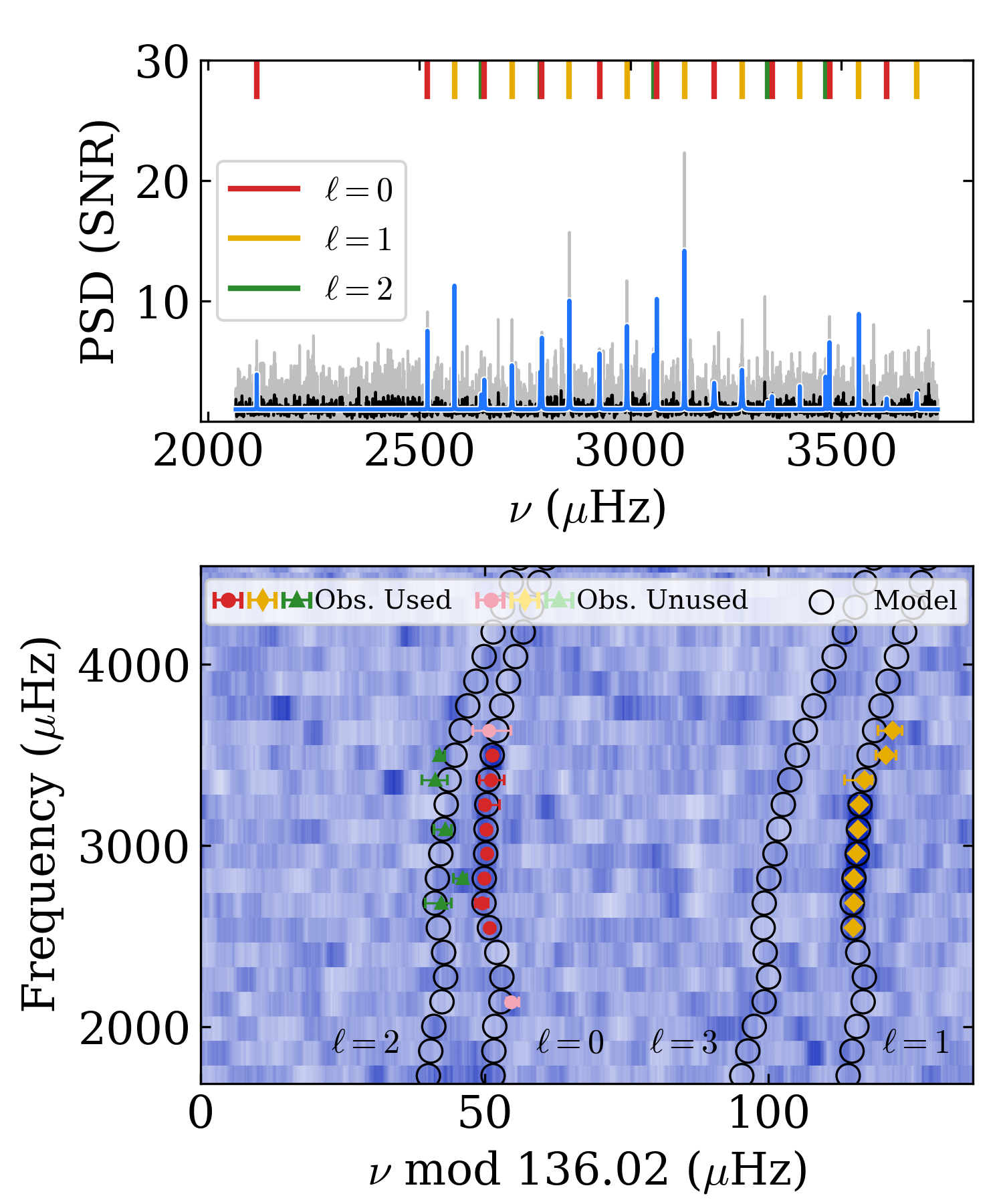}
    \caption{Same legend as in Figure~\ref{fig:echelle_3241581}; shown here for EPIC 206371648.}
    \label{fig:echelle_206371648}
\end{figure}

\subsection{EPIC 212624487}

\begin{table}[H]
\centering
\begin{tabular}{ccccc}
\hline
$n$ & $\ell$ & $\nu_{\texttt{apollinaire}}$ & $\nu_{\mathrm{PBJAM}}$ & extended \\
\hline
16 & 0 & 2313.43 $\pm$ 0.34 & 2311.85 $\pm$ 3.18 & 1 \\
17 & 0 & 2442.00 $\pm$ 3.84 & & 1 \\
18 & 0 & 2575.14 $\pm$ 0.93 & 2574.91 $\pm$ 1.84 & 0 \\
19 & 0 & 2707.26 $\pm$ 0.38 & 2707.04 $\pm$ 0.20 & 0 \\
20 & 0 & 2839.24 $\pm$ 0.31 & 2839.25 $\pm$ 0.24 & 0 \\
21 & 0 & 2972.30 $\pm$ 0.34 & 2972.40 $\pm$ 0.20 & 0 \\
22 & 0 & 3104.80 $\pm$ 0.29 & 3104.73 $\pm$ 0.23 & 0 \\
23 & 0 & 3236.80 $\pm$ 0.32 & 3236.82 $\pm$ 0.27 & 0 \\
24 & 0 & 3369.27 $\pm$ 1.04 & 3369.42 $\pm$ 1.02 & 0 \\
25 & 0 & 3505.02 $\pm$ 0.56 & 3505.06 $\pm$ 0.33 & 0 \\
26 & 0 & & 3638.00 $\pm$ 3.25 & 1 \\
27 & 0 & & 3771.19 $\pm$ 3.31 & 1 \\
15 & 1 & 2242.93 $\pm$ 1.17 & & 1 \\
16 & 1 & 2375.71 $\pm$ 1.01 & 2374.71 $\pm$ 1.46 & 0 \\
17 & 1 & 2504.67 $\pm$ 1.50 & 2504.36 $\pm$ 1.09 & 0 \\
18 & 1 & 2636.82 $\pm$ 0.31 & 2637.14 $\pm$ 0.29 & 0 \\
19 & 1 & 2769.19 $\pm$ 0.22 & 2769.22 $\pm$ 0.39 & 0 \\
20 & 1 & 2901.74 $\pm$ 0.30 & 2901.35 $\pm$ 0.33 & 0 \\
21 & 1 & 3034.65 $\pm$ 0.46 & 3034.72 $\pm$ 0.45 & 0 \\
22 & 1 & 3167.43 $\pm$ 0.37 & 3167.48 $\pm$ 0.44 & 0 \\
23 & 1 & 3299.97 $\pm$ 0.35 & 3300.10 $\pm$ 0.35 & 0 \\
24 & 1 & 3433.43 $\pm$ 1.07 & 3433.71 $\pm$ 0.86 & 0 \\
25 & 1 & 3566.52 $\pm$ 0.72 & 3566.48 $\pm$ 0.77 & 0 \\
26 & 1 & 3702.88 $\pm$ 2.90 & 3700.64 $\pm$ 2.15 & 0 \\
15 & 2 & 2298.83 $\pm$ 2.05 & & 1 \\
16 & 2 & 2431.60 $\pm$ 3.47 & & 1 \\
17 & 2 & 2562.35 $\pm$ 1.41 & 2563.25 $\pm$ 3.59 & 0 \\
18 & 2 & 2694.97 $\pm$ 0.53 & 2693.95 $\pm$ 0.83 & 0 \\
19 & 2 & 2826.11 $\pm$ 1.04 & 2827.09 $\pm$ 3.54 & 0 \\
20 & 2 & 2960.38 $\pm$ 0.98 & 2961.27 $\pm$ 0.83 & 0 \\
21 & 2 & 3094.01 $\pm$ 0.59 & 3093.27 $\pm$ 0.96 & 0 \\
22 & 2 & 3226.78 $\pm$ 0.91 & 3226.82 $\pm$ 1.71 & 0 \\
23 & 2 & 3356.93 $\pm$ 1.99 & 3357.37 $\pm$ 2.44 & 0 \\
24 & 2 & 3493.29 $\pm$ 1.72 & 3494.25 $\pm$ 1.26 & 0 \\
\hline
\end{tabular}
\caption{Same as in Table~\ref{tab:freq_3241581} shown here for EPIC 212624487.}
\label{tab:freq_212624487}
\end{table}
\vspace{-30pt}
\begin{figure}[H]
    \centering
    \includegraphics[width=0.90\linewidth]{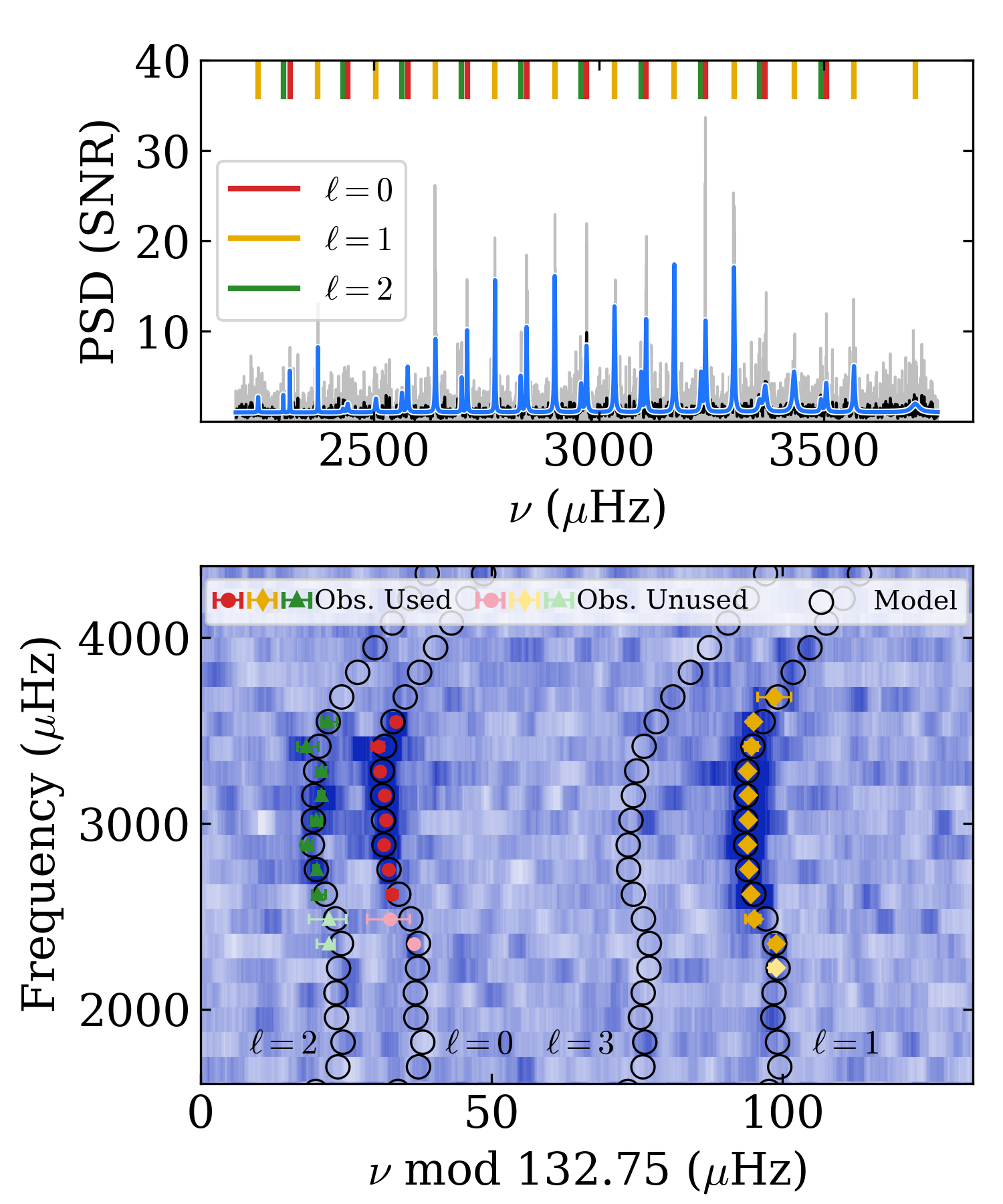}
    \caption{Same legend as in Figure~\ref{fig:echelle_3241581}; shown here for EPIC 212624487.}
    \label{fig:echelle_212624487}
\end{figure}

\subsection{EPIC 212708252}

\begin{table}[H]
\centering
\begin{tabular}{ccccc}
\hline
$n$ & $\ell$ & $\nu_{\texttt{apollinaire}}$ & $\nu_{\mathrm{PBJAM}}$ & extended \\
\hline
15 & 0 & 2209.99 $\pm$ 0.36 & 2210.06 $\pm$ 2.36 & 0 \\
16 & 0 & 2342.35 $\pm$ 0.26 & 2342.34 $\pm$ 1.25 & 0 \\
17 & 0 & 2473.76 $\pm$ 0.27 & 2473.44 $\pm$ 2.14 & 0 \\
18 & 0 & 2607.34 $\pm$ 0.15 & 2607.30 $\pm$ 0.15 & 0 \\
19 & 0 & 2740.69 $\pm$ 0.11 & 2740.69 $\pm$ 0.11 & 0 \\
20 & 0 & 2873.61 $\pm$ 0.12 & 2873.61 $\pm$ 0.11 & 0 \\
21 & 0 & 3007.39 $\pm$ 0.16 & 3007.37 $\pm$ 0.17 & 0 \\
22 & 0 & 3140.86 $\pm$ 0.16 & 3140.82 $\pm$ 0.19 & 0 \\
23 & 0 & 3274.48 $\pm$ 0.25 & 3274.47 $\pm$ 0.36 & 0 \\
24 & 0 & 3409.61 $\pm$ 0.61 & 3409.57 $\pm$ 1.18 & 0 \\
25 & 0 & 3547.70 $\pm$ 1.41 & 3546.49 $\pm$ 2.17 & 0 \\
14 & 1 & 2141.84 $\pm$ 2.19 & 2141.84 $\pm$ 2.06 & 0 \\
15 & 1 & 2272.46 $\pm$ 0.35 & 2272.55 $\pm$ 0.88 & 0 \\
16 & 1 & 2404.38 $\pm$ 0.21 & 2404.38 $\pm$ 0.23 & 0 \\
17 & 1 & 2537.33 $\pm$ 0.11 & 2537.45 $\pm$ 0.15 & 0 \\
18 & 1 & 2671.39 $\pm$ 0.17 & 2671.31 $\pm$ 0.18 & 0 \\
19 & 1 & 2804.66 $\pm$ 0.14 & 2804.68 $\pm$ 0.13 & 0 \\
20 & 1 & 2938.44 $\pm$ 0.10 & 2938.44 $\pm$ 0.10 & 0 \\
21 & 1 & 3072.33 $\pm$ 0.16 & 3072.36 $\pm$ 0.16 & 0 \\
22 & 1 & 3206.55 $\pm$ 0.21 & 3206.53 $\pm$ 0.23 & 0 \\
23 & 1 & 3340.98 $\pm$ 0.21 & 3340.97 $\pm$ 0.24 & 0 \\
24 & 1 & 3475.33 $\pm$ 0.48 & 3475.32 $\pm$ 0.50 & 0 \\
25 & 1 & 3609.93 $\pm$ 1.58 & 3609.13 $\pm$ 2.59 & 0 \\
15 & 2 & 2334.17 $\pm$ 0.16 & 2334.19 $\pm$ 0.70 & 0 \\
16 & 2 & 2466.03 $\pm$ 0.90 & 2466.70 $\pm$ 2.26 & 0 \\
17 & 2 & 2599.66 $\pm$ 0.39 & 2599.42 $\pm$ 0.47 & 0 \\
18 & 2 & 2733.18 $\pm$ 0.88 & 2733.35 $\pm$ 0.66 & 0 \\
19 & 2 & 2867.38 $\pm$ 0.14 & 2867.42 $\pm$ 0.14 & 0 \\
20 & 2 & 3001.63 $\pm$ 0.19 & 3001.63 $\pm$ 0.19 & 0 \\
21 & 2 & 3135.27 $\pm$ 0.28 & 3132.06 $\pm$ 1.96 & 0 \\
22 & 2 & 3269.95 $\pm$ 0.28 & 3269.72 $\pm$ 1.54 & 0 \\
23 & 2 & 3402.64 $\pm$ 2.37 & & 1 \\
24 & 2 & 3539.62 $\pm$ 2.83 & 3537.22 $\pm$ 4.00 & 1 \\
\hline
\end{tabular}
\caption{Same as in Table~\ref{tab:freq_3241581} shown here for EPIC 212708252.}
\label{tab:freq_212708252}
\end{table}
\vspace{-30pt}
\begin{figure}[H]
    \centering
    \includegraphics[width=0.9\linewidth]{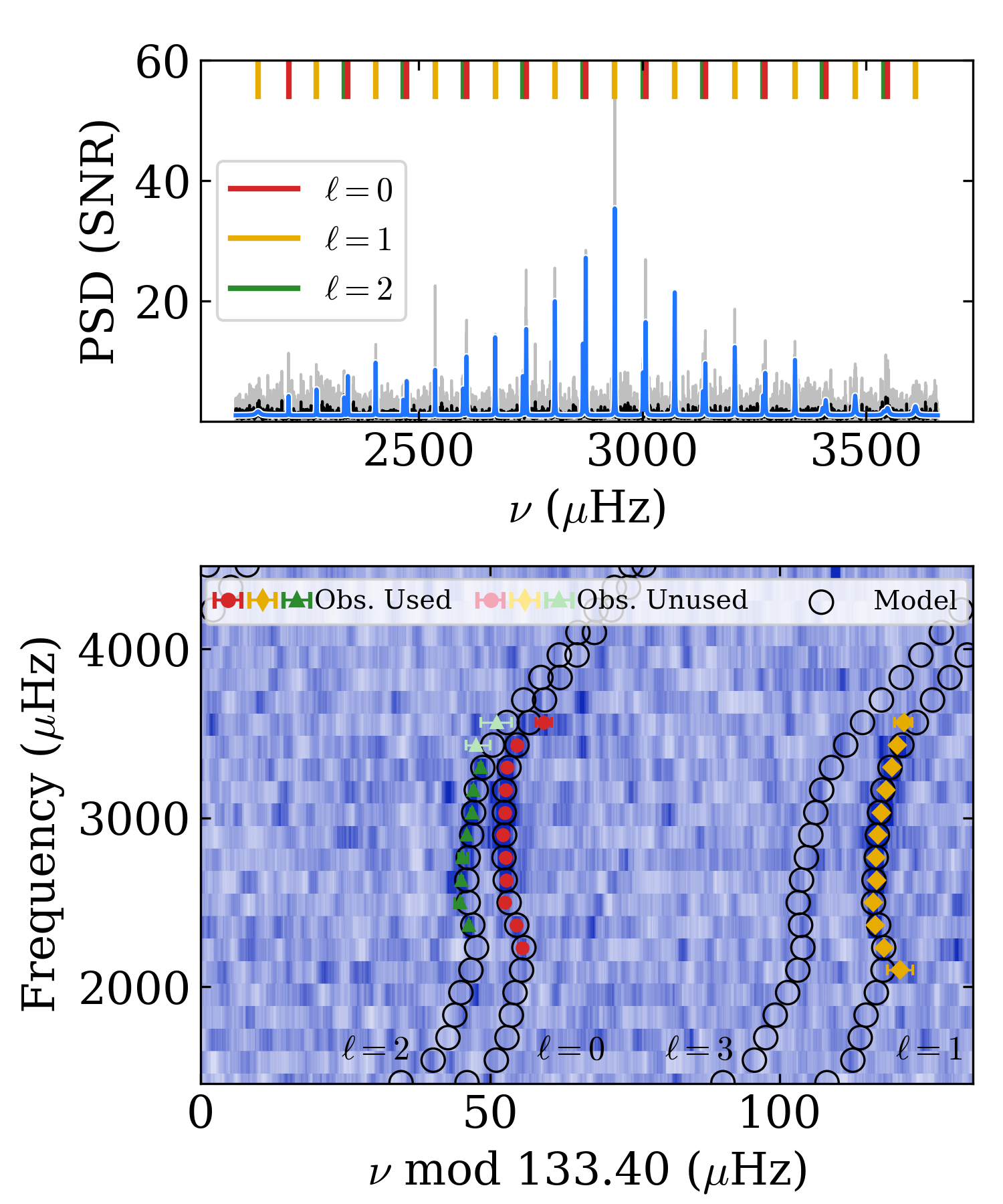}
    \caption{Same legend as in Figure~\ref{fig:echelle_3241581}; shown here for EPIC 212708252.}
    \label{fig:echelle_212708252}
\end{figure}


\newpage
\twocolumn
\section{Radial velocity analysis of KIC\,3241581}
\label{app:radial_velocity}

\begin{figure}[h]
\begin{center}
    \includegraphics[width=\columnwidth,trim={0cm 0.5cm 0cm 0.5}]{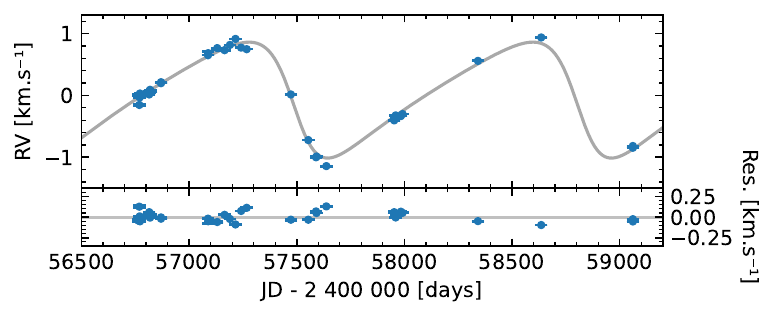}
\end{center}    
    \caption{Spectroscopic observations and orbital model  of KIC\,3241581 from the Mercator/HERMES data (top panel), with the residuals shown in the bottom panel.} 
    \label{fig:orbit_3241581}
\end{figure}

KIC\,3241581 was reported to be a binary by \cite{Beck2016} and has been continuously monitored for more than six years with the same instrumentation. We fitted a Keplerian orbit to the 40 available radial-velocity (RV) measurements, obtaining a period of $1316.4 \pm 1.0$\,days and a moderate eccentricity of $0.367 \pm 0.002$. The full set of orbital parameters is listed in Table\,\ref{tab:orbit_3241581}. The RV time series together with the best-fit model and the corresponding residuals are shown in Fig.\,\ref{fig:orbit_3241581}.

The peak-to-peak RV amplitude of $\sim2$\,km\,s$^{-1}$ is relatively small compared to samples of other main-sequence binaries \citep[e.g.][]{Beck2017,2026A&A...707A.298B}. The combination of a high RUWE value ($\sim6$) and small RV variations suggests that the orientation of the system may be close to plane-on.

\begin{table}[h]
\centering
\small
\caption{Orbital elements of KIC\,3241581.\label{tab:orbit_3241581}}
\begin{tabular}{lr}
\hline
\noalign{\smallskip}
 Parameter & Value \\
 \hline
  \noalign{\smallskip}
$P$ {[}d{]} & 1316.4\,$\pm$\,1.0  \\
$e$ & 0.367\,$\pm$\,0.002  \\
$T_0$ {[}d{]} & 2\,457\,500.6\,$\pm$\,1.2  \\
$K$ {[}km/s{]} & 0.940\,$\pm$\,0.002  \\
$\omega$ {[}\textdegree{]} & 103.3\,$\pm$\,0.3  \\
$\gamma$ {[}km/s{]} & -30.966\,$\pm$\,0.001 \\
$f$(m) {[}M$_\odot${]} &  0.000091\,$\pm$\,0.000001\\
O-C {[}km/s{]} &  0.05\\
\noalign{\smallskip}
\hline
\end{tabular}
\end{table}

\end{appendix}

\end{document}